\documentclass[useAMS,usenatbib]{mn2e}

\usepackage{graphicx}

\title[]{The Tully-Fisher relation of intermediate redshift 
field and cluster galaxies from Subaru spectroscopy
\thanks{Based on data collected at Subaru Telescope, which is operated by the National Astronomical Observatory of Japan.}}
\author[]{O. Nakamura$^{1}$\thanks{E-mail:osamu.nakamura@nottingham.ac.uk},
          A. Arag{\' o}n-Salamanca$^{1}$,
          B. Milvang-Jensen$^{2}$,
          N. Arimoto$^{3}$,
\newauthor
          C. Ikuta$^{3}$,
          S. P. Bamford$^{1}$\\
$^{1}$School of Physics and Astronomy, University of Nottingham,
      University Park, Nottingham, NG7 2RD, UK\\
$^{2}$Max-Planck-Institut f{\" u}r extraterrestrische Physik,
      Giessenbachstra\ss e, 85748 Garching, Germany\\
$^{3}$National Astronomical Observatory, 2-21-1 Osawa, Mitaka,
      Tokyo 181-8588, Japan\\
}
\begin{document}

\date{Accepted ???. Received ???; in original form ???}

\pagerange{\pageref{firstpage}--\pageref{lastpage}} \pubyear{2005}

\maketitle

\label{firstpage}

\begin{abstract}
We have carried out spectroscopic observations
in 4 cluster fields using Subaru's FOCAS multi-slit spectrograph 
and obtained spectra for 103 bright disk field and cluster 
galaxies at $0.06 \le z \le 1.20$. Seventy-seven of these show
emission lines, and 33 provide reasonably-secure
determinations of the galaxies' rotation
velocity.
The rotation velocities, luminosities, colours and emission-line properties of
these galaxies are used to study the possible effects of the cluster
environment on the star-formation history of the galaxies. 
Comparing the Tully-Fisher relations of cluster and field galaxies
at similar reshifts we find no measurable difference in rest-frame 
$B$-band luminosity at a given rotation velocity 
(the formal difference is $0.18\pm0.33\,$mag). 
The colours of the cluster emission line galaxies are only marginally redder 
in rest-frame $B-V$ (by $0.06\pm0.04\,$mag) 
than the field galaxies in our sample.
Taken at face value, these results seem to indicate
that bright star-forming cluster spirals 
are similar to their field counterparts in their star-formation properties. 
However, we find that the fraction of disk galaxies with absorption-line 
spectra (i.e., with no current star formation) is larger in clusters than in
the field by a factor of $\sim3$--$5$. This suggests  
that the cluster environment 
has the overall effect of switching off star formation in 
(at least) some spiral galaxies. 
To interpret these observational results, we carry out  simulations of the
possible effects of the cluster environment on the star-formation history of
disk galaxies and thus their photometric and spectroscopic properties.  This
allows us to create mock samples of unperturbed ``field'' galaxies (with
approximately constant star-formation rates) and perturbed ``cluster'' galaxies
with  different star-formation histories, including star formation truncation, 
with or without an associated starburst.  We show that, if we select only
bright galaxies with current star-formation (i.e,, with emission lines strong
enough for rotation-curve measurements), the average  colours and luminosities
of the ``cluster'' galaxies may not be very different from those  of galaxies
in the ``field'' sample,  even though their  star-formation histories may be
significantly different.  However, the fraction of emission and absorption-line
galaxies would change significantly. We also use these simulations to 
estimate the size of field and cluster galaxy samples that would allow us to 
differentiate the different star-formation scenarios considered. 
Finally, we find that the rest-frame absolute $B$-band magnitude of
the field galaxies in our sample shows an evolution of
$-1.30\pm1.04$ mag per unit redshift at fixed rotation velocity.
This indicates that the average
SFR of bright disk galaxies evolves more slowly than 
the universal star-formation rate as determined from 
UV, H$\alpha$, far-infrared and radio
studies. This suggests the evolution of the universal 
SFR density is not dominated by bright star-forming disk galaxies,
in agreement with previous studies. 
\end{abstract}


\section{Introduction}

Recent studies of distant galaxies have established that
the fraction of S0s in rich galaxy clusters drops by a factor 2--3
from the local universe to $z\sim0.5$, while the fraction of
spiral galaxies increases comparatively  (Couch et al. 1994;
Dressler et al. 1997; van Dokkum et al. 1998; Fasano et al. 2000).
Parallel studies have found a number of poststarburst galaxies within
clusters (Dressler \& Gunn 1983; Couch \& Sharples 1987),
a suppressed star formation rate for spiral types relative to the
field (Balogh et al. 1998), and evidence of an infalling
spiral population out to $z\sim0.4$ (Poggianti et al. 1999;
Kodama \& Bower 2001). This naturally leads to the scenario
that cluster S0s could form from spiral galaxies through interaction
with the cluster environment (Jones, Smail \& Couch 2000;
Kodama \& Smail 2001).

Three mechanisms have been suggested for the transformation: galaxy-galaxy
interactions, tidal forces, and gas stripping. Galaxy-galaxy interaction is
most efficient in group environments, where relative velocities are low
(Zabludoff \& Mulchaey 1998; Mulchaey \& Zabludoff 1998; Ghigna et al. 1998),
while  tidal forces (Toomre \& Toomre 1972; Moore et al. 1996; Mihos, McGaugh,
\& de Blok 1997) are effective on small spirals. Gas stripping is expected to
be an effective process in  rich clusters (Abadi, Moore, \& Bower 1999;
Quilis, Moore, \& Bower 2000; Bekki, Couch, \& Shioya 2002; Vogt et al. 2004).
Although each mechanism is efficient in different
circumstances and/or environments, many of the 
numerical studies of these mechanisms predict
a star-burst could/should happen during the processes leading to 
the cessation of star formation required by the spiral-to-S0 transformation. 

Until recently, however, there has been no compelling evidence indicating 
that spiral galaxies falling into dense environments experience a  star-burst.
Motivated by this  Ziegler et al. (2003) investigated three clusters at
$z=0.3$--$0.5$ and used the Tully-Fisher relation (TFR)  to compare galaxy
luminosities of field and cluster spirals at  the same mass, finding no
measurable difference.  In contrast, Milvang-Jensen et al. (2003) carried out a
similar study  for the rich cluster MS1054.4$-$0321 at $z=0.83$, and found
evidence for some brightening in the cluster galaxies relative to the field
ones at the same rotation velocity, implying more active star-formation of
spiral galaxies in clusters. The situation is thus controversial, but these
studies showed that using the TFR as a tool for comparing field and
cluster galaxies at intermediate redshift could be successful.

Following Milvang-Jensen et al., we have carried out a study of the TFR in nine
additional distant rich clusters. Five have been observed with the VLT using
the FORS2 multi-slit spectrograph (Seifert et al. 2000)  and the results
(including a re-analysis of the  MS1054.4$-$0321 data) have been reported in
Bamford et al. (2005A, B05A hereafter), where they have found further evidence
for the brightening in the cluster spirals.  In this paper, we report on our
TFR study for the remaining four clusters observed with the Subaru Telescope
using the FOCAS spectrograph (Kashikawa et al. 2002). Since our sample includes
field galaxies, we also discuss their evolution following Bamford,
Arag\'on-Salamanca \& Milvang-Jensen  (2005B, B05B hereafter).

Before comparing the cluster spiral galaxies with the field ones, one needs to
be aware that the comparison could be complicated by strong selection effects. 
For example, the probability of observing a galaxy in a star-burst phase will
be lower if  the star-formation occurs over a short time-scale; on the other
hand, a star-bursting  galaxy will be brighter, and thus more likely to  appear
in a magnitude limited sample. Similarly, one could think that if our sample
contains galaxies  in the fading phase after the starburst has ended, the
average luminosity of our sample would be lowered, perhaps cancelling the
effect of the  initial brightening. However, it is unlikely that we will be
able to measure rotation curves for galaxies after the cessation of star
formation since  emission lines would be absent. All these effects will
probably affect the average properties of the galaxies in our field and cluster
samples, so we have carried out  careful modeling for the correct
interpretation of the observations.

Throughout the paper, we assume $\Omega_{m}=0.3$,
$\Omega_{\Lambda}=0.7$, and $H_0$=70 km s$^{-1}\,$Mpc$^{-1}$.
All magnitudes are in the Vega zero-point system.

\section[]{Data}

\subsection{Observations and data reduction}
\label{s_obs}

\begin{table*}
 \begin{minipage}{180mm}
  \caption{Summary of the clusters and spectroscopic observations.}
  \begin{tabular}{llllrcrrcccc}
  \hline
  Cluster &  \multicolumn{1}{c}{R.A.}
          &  \multicolumn{1}{c}{Dec.}
          &  \multicolumn{1}{c}{z}
          &  \multicolumn{1}{c}{$\sigma$}
          & Mask
          & P.A.
          & \multicolumn{1}{c}{$T_{\rm exp}$}
          & Seeing
          & \multicolumn{1}{c}{$N_{\rm slit}$}
          & \multicolumn{1}{c}{$N_{\rm gal}$}
          & \multicolumn{1}{c}{$N_{\rm dup}$$^c$} \\
          & \multicolumn{1}{c}{(J2000)}
          & \multicolumn{1}{c}{(J2000)}
          &
          &  \multicolumn{1}{c}{(km s$^{-1}$)}
          & ID
          & (deg)
          & \multicolumn{1}{c}{(s)}
          & (arcsec)
          &
          &
          & \\
  \hline
  A2390$^a$           & 21 53 36.8 & $+$17 41 32 & 0.2282 & 1294 & P1 &   $0\degr$ & $1200\times4$ & 0.7 &  12 &  14 & 1\\
                      &            &             &        &      & P2 & $-45\degr$ & $1200\times3$ & 0.5 &  13 &  13 &  \\
  MS1621.5$+$2640$^a$ & 16 23 34.5 & $+$26 34 17 & 0.4271 &  735 & P1 &  $45\degr$ & $1800\times4$ & 0.4 &  15 &  17 & 1\\
                      &            &             &        &      & P2 & $-45\degr$ & $1800\times4$ & 0.6 &  15 &  16 &  \\
  MS0015.9$+$1609$^a$ & 00 18 33.5 & $+$16 26 03 & 0.5490 &  984 & P1 &   $0\degr$ & $1800\times5$ & 0.5 &  16 &  16 & 2\\
                      &            &             &        &      & P2 &  $90\degr$ & $1800\times5$ & 0.5 &  20 &  22 &  \\
  MS2053.7$-$0449$^b$ & 20 53 44.6 & $-$04 49 16 & 0.583  &  817 & P1 &   $0\degr$ & $1800\times5$ & 0.4 &  17 &  18 &  \\
  \hline
  \end{tabular}
  \label{t_cls_obs}

\medskip
$^a$ Position, z and $\sigma$ from Girardi \& Mezzetti (2001).\\
$^b$ Position and z from Stocke et al. (1991),
                     $\sigma$ from Hoekstra et al. (2002).\\
$^c$ Number of galaxies observed in two masks.
\end{minipage}
\end{table*}

Our sample contains rich clusters, mainly in the northern hemisphere, 
covering a wide redshift range, similar to B05A.
Table~\ref{t_cls_obs} lists the clusters observed, together 
with some of their properties and basic observational 
parameters. The MS2053.7$-$0449 cluster was also studied by 
B05A, so we should be able to test the 
consistency of the rotation velocity measurements.

Pre-imaging in the $R$-band was carried out on 2002 June 7 using 
FOCAS in imaging mode. The field-of-view was
6 arcmin diameter, and the pixel scale was 0.1 arcsec pixel$^{-1}$.
Each field was taken at two different position angles to
enable flexible allocation of tilted slits (see below). The exposure
times at each position angle were 180s for A2390 and 240s for
the other clusters. The seeing was 0.5--0.6 arcsec. The images at the
different position angles were combined,
and SExtractor version 2.0 (Bertin \& Arnouts 1996)
was used to identify the objects.
The photometry was calibrated to Thuan \& Gunn (1976) $r$-band
(see Sec.\ref{s_absmag}).

For multi-object spectroscopy (MOS) 
targeting, we determined the priority of each galaxy according to how
many of the following conditions were met:  (i) disky morphology in pre-imaging
(must), (ii) bright enough to observe, (iii) inclination $\ga$ 45$\degr$, (iv)
HST images available from the archive, (v) magnitudes and/or the redshift
available from literature, and (vi) known to be a cluster member. We designed
the masks by allocating slits to galaxies with the highest priorities. The
slits were to be placed along the major axis of galaxies to measure rotation
velocities. Two masks with  complementary position angles were designed for
each cluster except MS2053.7$-$0449 for which we decided to use only one mask.
This allowed us to use slit angles   $\la 45\degr$ relative to the
spatial direction for most galaxies, avoiding degrading the resolution
beyond acceptable limits. A few slits included more than one galaxy by chance.
Table~\ref{t_cls_obs} gives the details of the masks. In total 108 slits were
placed on 116 galaxies in 7 masks, of which 4 galaxies were observed in two
masks and 112 were unique. In the following we always refer to the number of
unique galaxies unless stated. The morphological distribution of the 
selected galaxies is discussed in Sec.~\ref{s_morph}. 
The procedures followed for target
selection, priority allocation and mask design were very similar to those used
by B05A.

The spectroscopic observations were carried out on 2002 Aug 10--11 using FOCAS
in MOS mode. The 300B grism was used, yielding  a dispersion of
1.4{\AA}$\,$pixel$^{-1}$. The slit width was  0.6 arcsec along the dispersion
axis for all the objects, achieving a spectral resolution 
$R \simeq 1200$, slightly higher than
B05A. The lower limit of the spectral range was set to 4700 {\AA} by the order
blocking filter, and the upper limit was up to 9400 {\AA}, depending on the
geometrical position of each slit on the mask. The seeing, determined from
the spectra of $\sim2$ stars observed in each mask, was $0.4$--$0.7$
arcsec, and we binned $\times 2$ along the spatial direction, achieving a
final  spatial sampling of 0.2 arcsec pixel$^{-1}$ for untilted
slits. Given the differences in
redshift, we  made shorter exposures for A2390 and longer ones for
MS2053.7$-$0449 to achieve roughly the same depth in every cluster.
Table~\ref{t_cls_obs} gives the exposure time of each mask.

Bias subtraction, flat fielding, connection of the two CCD detectors, and
distortion correction were done using the FOCASRED package, developed
by the instrument team. Wavelength calibration and sky-subtraction
were carried out with standard IRAF tasks to obtain the 2-dimensional spectra.

\subsection{Redshift and cluster membership}
\label{s_z}

The centre of each galaxy on the 2D spectrum was identified by the peak
of the continuum component without emission lines. Central
spectra were extracted for each galaxies
in $0.2$--$0.6$ arcsec apertures,
and the redshift was measured using emission and/or absorption
features. When more than one emission line were well detected,
the mean of the redshifts was taken.
The emission lines included
[O II]$\lambda3727$, H$\delta$, H$\gamma$, H$\beta$,
[O III]$\lambda4959$, [O III]$\lambda5007$, [N II]$\lambda6548$,
H$\alpha$, [N II]$\lambda6583$, [S II]$\lambda6716$, and
[S II]$\lambda6731$. The redshifts of 77 galaxies were measured
in this way.
When emission features were not seen, absorption features
were used instead. We employed a cross-correlation
method with a model template spectrum. Redshifts for 27 absorption-line
galaxies were
measured in this way. We could not determine the redshifts
of 8 galaxies because of poor detection of their absorption features or 
dubious 
emission line identification. We drop these galaxies from
further analysis, and the remaining is 104. The redshift range
of the galaxies was $0.06 \le z \le 1.20$ with median of $z=0.42$.
When considering only 
the galaxies in the final TFR sample (see Sec.~\ref{s_rtv}), the 
redshift range was 
$0.06 \le z \le 0.74$, with a median of $z=0.39$.

The field-of-view of FOCAS was too small to cover 
the whole of the clusters, so 
cluster membership was decided on the basis of velocity only, without 
considering the spatial location of the galaxies. 
Galaxies with velocities within 
$\pm3 \sigma$ from the velocity centre of the cluster
were classified as cluster members. The 
fraction of cluster and field galaxies were 48$\%$ and 52$\%$,
respectively. The numbers for each cluster are summarised in
Table~\ref{t_nem}. 

\subsection{Absolute $B$-band magnitudes and inclinations}
\label{s_absmag}

To supplement the pre-imaging photometry and obtain colour information, we
obtained  magnitudes for the targeted galaxies from the literature.
We also collected WFPC2 images from the archive, as in B05A.
The available sources of photometry, 
other than our own FOCAS pre-imaging, were the following: 
for A2390, the CNOC survey  (Yee et al. 1996) and 
                                        WFPC2 images  
					($F555W$, $F702W$, $F814W$);
for MS1621.5$+$2640, CNOC  (Ellingson et al. 1997) plus
                                        WFPC2 ($F555W$, $F814W$);
for MS0015.9$+$1609, CNOC (Ellingson et al. 1998) plus
               data from Dressler \& Gunn (1992), and WFPC2 ($F555W$, $F814W$);
for MS2053.7$-$0449,                        WFPC2 images 
($F606W$, $F702W$, $F814W$).
Note that not all 
these data were available for all of our targeted galaxies (see below).

The CNOC survey used the Thuan \& Gunn photometric system and 
provides $r$-band total magnitudes and $g-r$ colours within
6.4 arcsec apertures.
Dressler \& Gunn (1992) used Schneider, Gunn, \& Hoessel (1983)
photometric system, and give total $g$-band magnitudes
and $g-r$ and $r-i$ colours within the fitting radius.
The WFPC2 images were extracted from the archive with the basic
reductions completed. We ran SExtractor for these images
and obtained AUTOMAGs, which provide good estimates
of total magnitudes within 0.1 mag. This uncertainty in 
the total magnitude is too small to have any effect on our TFR analysis.
The zero-points were calibrated with reference to the latest
WFPC2 manual.
The available HST photometric bands differed from 
cluster to cluster and object to object,
but at least two bands were always available, so we had some colour
information for every galaxy within the WFPC2 field. 
The redder bands were used to define the aperture centres when deriving
colours.
The FOCAS pre-imaging was not photometrically-calibrated at the telescope, 
so that photometry
was calibrated using the galaxies that were in common with the
CNOC. A constant offset was used to transform SExtractor's AUTOMAG 
in the pre-imaging to the total Thuan \& Gunn $r$-band magnitudes from CNOC.
The error of this calibration was $\sim0.08$ mag, estimated from the
$r.m.s.$ of the transformation. 
Thus, the magnitudes derived from the
FOCAS pre-imaging are in the Thuan \& Gunn $r$-band
system. 

For each galaxy, the magnitude in the band that was closest
to the rest-frame Johnson-Morgan $B$-band was chosen among those
available as the starting
point to calculate rest-frame $B$ magnitudes.  
To obtain a  representative colour we used the data from 
the CNOC, WFPC2, and Dressler \& Gunn (1992) (in that order of 
priority). The Galactic extinction of
Schlegel, Finkbeiner, \& Davis (1998) was applied to
these magnitudes and colours.
The magnitude was then converted to the absolute rest-frame $B$-band
magnitude by refering to the galaxy colours and
K-corrections of Fukugita, Shimasaku, \& Ichikawa (1995).
The concordance cosmology (cf. \S1) 
was used to calculate the distance modulus.
In many cases we had several colours available from different 
sources, allowing us to estimate
that the error in the rest-frame $B$-band magnitude associated with using
different colours was 0.03 mag on average.
Moreover, we estimate that the error associated with using different 
photometric bands to estimate 
rest-frame $B$-band was 0.08 mag on average.
This indicates that the values of our rest-frame $B$ magnitudes are
quite robust, with uncertainties of the order of 0.1 mag or less. 

In a few cases the only available photometry  came from the FOCAS
pre-imaging. In that case, the rest-frame $B$ magnitude  was estimated from the
pre-imaging magnitude applying an Sc galaxy K-correction. Nine objects 
belong to this category, with only 
four making it to the final TFR study.

The inclination of the disk component was measured using 
GIM2D (Simard et al. 2002) following B05A.
The WFPC2 images were used whenever available, while
the pre-imaging was used otherwise.
When the WFPC2 images were used, we always selected the band
with the highest S/N, yielding the highest confidence.
There were 48 galaxies with an inclination $i>40\degr$
(i.e., large enough for reliable rotation velocity determination)
which had their inclinations measured 
in both the pre-imaging 
and the WFPC2 $F814W$ images. 
For these galaxies, the average ratio of the $(\sin i)^{-1}$ values determined 
from the pre-imaging and HST data respectively was $1.02\pm0.01$.
Similarly, there were 39 galaxies with $i>40\degr$ that had their inclinations
determined from  $F555W$ and $F814W$. For these, the average 
$(\sin i)^{-1}$ ratio was $0.98\pm0.01$.
Hence the systematic error in mixing the inclinations obtained 
from the different WFPC2 bands
and the pre-imaging is only $\sim2\%$, and can be safely ignored.
Note that the FOCAS images had very good seeing, and the value of the 
inclination
is dominated by the outer isophotes, making our ground-based data perfectly
adequate for inclination determination at these moderate redshifts. 

\begin{figure}
   \includegraphics[width=84mm]{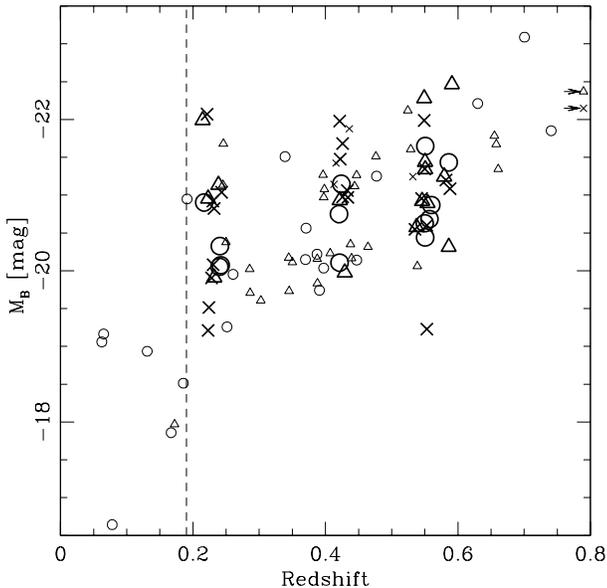}
   \caption{Absolute $B$-band magnitude vs. redshift
            for the 104 galaxies with measured redshifts.
            The different symbols indicate 
	    absorption-line galaxies (crosses), 
	    emission-line galaxies (triangles),
            and emission-line galaxies with measured rotation velocity
            (circles). Small and large symbols denote
            field and cluster galaxies, respectively. The two
            marks with arrows are the galaxies at $z=1.195$ (cross)
            and $1.041$ (triangle).
	    The dashed line at $z=0.19$
            is used to define an unbiased sample for our analysis
	    (see text for details).
}
   \label{f_z_B}
\end{figure}

Finally, the internal reddening was corrected as a function of
inclination following the prescription of Tully \& Fouque (1985).
The absolute extinction at face-on position was assumed to be 0.27
mag in $B$ following B05A.
The derived absolute $B$-band magnitudes ($M_{B}$) are shown in
Fig.~\ref{f_z_B} as a function of redshift.

There were seven galaxies in common with the VLT sample of B05A
in the MS2053.7$-$0449 field. Our $B$-band magnitudes for these
agreed well with those by B05A, with five out of the seven being within
0.1 mag. This ensures the consistency of the derived magnitudes even though
we used partly different procedures and photometric sets
from B05A in our derivation.

Our sample does not contain enough bright galaxies at the lower end of the
redshift range due the small observed volume. Since this could cause a bias in
our analysis, we will only consider  galaxies with $z>0.19$ in our study. In
doing so we ensure that our field and cluster samples contain galaxies covering
similar luminosity ranges ($-22.5 \la M_{B} \la -19.5$).  The redshift and
magnitude ranges studied here are comparable to the 'matched' sample of B05A
($z>0.25$ and $M_{B}<-19.5$), although we do not reach the 
redshift of their most distant 
cluster (MS1054.4$-$0321 at $z=0.83$). 

\subsection{Rotation velocities}
\label{s_rtv}

In order to be able to measure the rotation velocities of the 77 galaxies with
detected emission lines (see Table~\ref{t_nem}), we first fitted the continuum
component and removed it in each spectrum by interpolation.  Then every
emission line was cut out to a postage stamp image, and input to the software
ELFIT2PY (B05A) together with the inclination and the absolute $B$-band
magnitude to measure the rotation velocity. ELFIT2PY is based on the
algorithm of ELFIT2D
(Simard \& Pritchet 1998, 1999), and calculates the best fitting the
rotation velocity and scale-length of the emission assuming an intrinsic
rotation curve and an exponential
light  distribution. The software takes into account the instrumental
resolution and the seeing,  as well as the slit width 
(optical beam smearing).  The output rotation velocity has to be corrected
for inclination. The inclination is also  necessary as an input to calculate
the light distribution through  the slit. The absolute $B$-band
magnitude is taken into account  in the fitting process since we assumed the
universal rotation curve of Persic \& Salucci (1991). The output 
rotation velocity would change, on average, by $\sim 10$ km s$^{-1}$ 
if we
adopted a flat rotation curve, the exact value depending on 
the luminosity/mass of the galaxy (cf.\ B05A). Such small
change would not affect the
conclusions of this study.
The spatial
centre was determined from the centroid of the continuum, while we allowed 
$\pm 1$ pixel flexibility in the velocity centre ($\sim
76$ km s$^{-1}$ at 5500 {\AA} and $z = 0$) to let ELFIT2PY account for
possible errors in redshift
and wavelength calibration. In fact, of the objects used in the final
TFR (see below) 85\% had offsets of 0.2 pixels or less.

As part of the rotation velocity determination we carried out  a
quality-control process to ensure that only reliable measurements were used in
the analysis. Our process is similar to the one used by B05A, so we
will only describe here briefly the main quality-control steps. First, we
did not attempt to measure the rotation velocities  when the emission was
nuclear (i.e., spatially-unresolved) or when  the mean $S/N$ was below a
given threshold.
Second, when measurement yielded a rotation velocity smaller than
its estimated $1\sigma$ error, we rejected the measurement as unreliable.
Third, the fits were visually inspected and those clearly wrong were also
rejected. The  main causes for rejection included: not reaching the flat
rotation or the turn-over point (cf. Verheijen 2001), clear
non-exponential distribution of the emission, strong asymmetry,
and a deep stellar absorption close to the emission.
In a few cases the slits appeared to have been cut in the 
wrong direction, so these galaxies were also rejected.
Note that our treatment of the turn-over point
is different from that of B05A, and we will discuss this issue
in Sec.~\ref{s_comp}. 
After this process, only 33 galaxies (43$\%$) with detected
emission lines yielded secure rotation velocities
that we can reliably use in the final TFR analysis. 
For reference, we attempted to fit 63 galaxies using ELFIT2PY,
of which 15 were rejected due to clear velocity under-estimates by 
ELFIT2PY or no sign of turn-over,
9 for low quality fits, 2 for lack of rotation structure, and
4 for problems not directly related to the 
rotation velocity measurement.

When more than one emission line were securely measured for a galaxy,
we used the velocity measured using the line with the highest S/N, 
while B05A took the error-weighted mean value. 
The number of emission lines measured per galaxy was 2.5 on average
for the 33 galaxies, and 25 of these had more than one emission line.
The ratio of the rotation velocities determined with the
two highest S/N lines has a mean of  
$0.97$ and a standard deviation of $0.11$.
Hence the rotation velocities of the different emission lines
were consistent with each other typically within $11\%$.

\begin{figure*}
 \centering
 \begin{minipage}{170mm}
   \includegraphics[width=140mm]{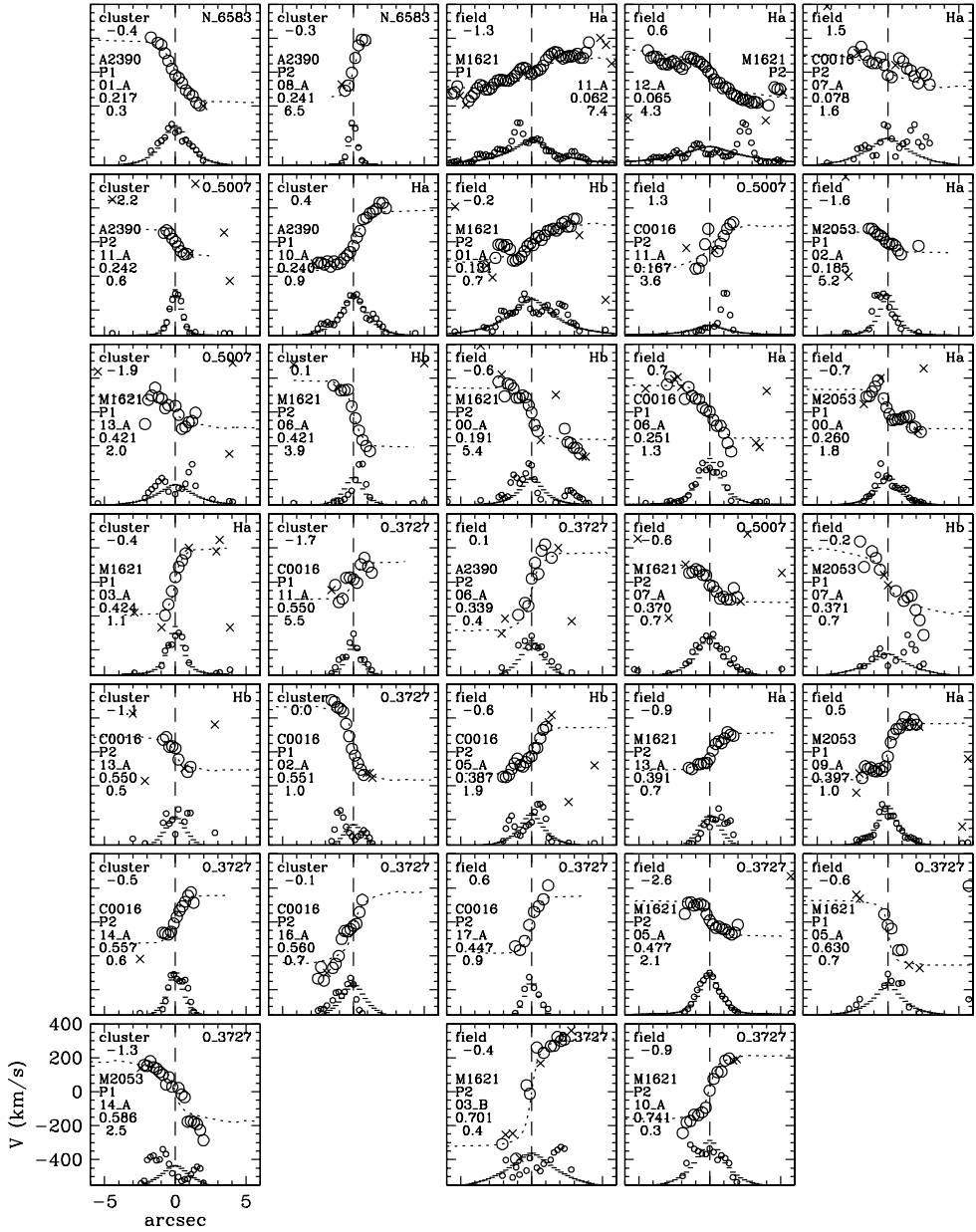}
   \caption{The observed rotation curves of the 33 galaxies that survive 
            our quality-control process and are used in the
            final TFR analysis. 
	    The cluster and field galaxies are displayed
            in the first two and the last three columns, respectively,
            in redshift order.
            The velocities are ``observed'', as opposed to ``intrinsic'', i.e., 
	    not corrected for inclination, 
	    non-zero slit width, seeing effects, etc. 
            The vertical dash line shows the spatial centre.
            The large circles denote the velocity centre determined
	    from  Gaussian
            fits to each spatial column with the peak higher than
            2 times the background RMS and the FWHM between 0.7
            and 1.4 the resolution. The crosses are the same as
            the circles but with the peak as low as
            0.7 times the background RMS
            and the FWHM between 0.5 and 2.0.
            Dotted lines show the ELFIT2PY fitted model
            when the peak of the Gaussian is above 0.01 times
            the background RMS. The zero-point of the velocity
            in each panel is set to the adopted redshift for the object.
            The small circles and horizontal ticks at the bottom
	    of each panel show the flux
            distribution of the spectrum and model fit, respectively.
            The Flux is derived by integrating under the
            fitted Gaussians. 
            One pixel corresponds to $0.2 (\cos\theta)^{-1}$ arcsec
            where $\theta$ is the slit tilt angle. The width of the
            box is $12.2$ arcsec.
            Labels at the top-left of
            each panel indicate whether the galaxies belong to the clusters
	    or the field. 
            The numbers right below these labels 
	    indicate the residuals, in $M_B$, from the local TFR 
	    (cf. Fig.~\ref{f_zev}; see Sec.~\ref{s_tfr}).
            The cluster name, mask ID, slit ID, redshift,
            and $\chi^2$ per fitting element
            of the galaxies are indicated on the middle-left of each panel. 
            The emission line ID is shown on the top-right corner.
   }
   \label{f_rtv}
  \end{minipage}
\end{figure*}

\begin{figure}
   \includegraphics[width=84mm]{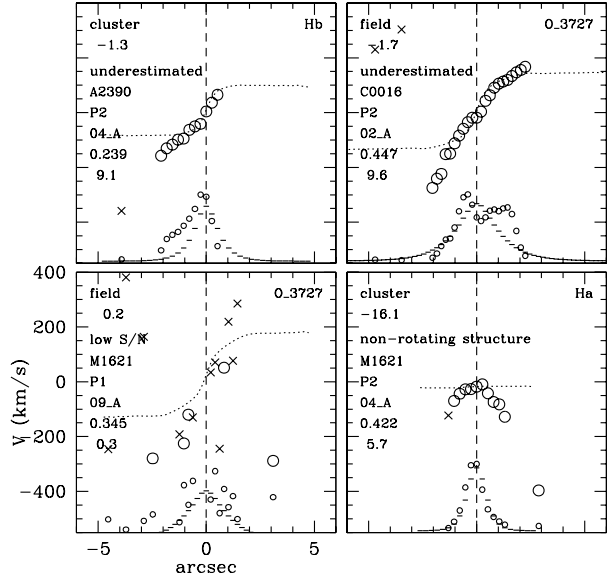}
   \caption{Examples of rotation velocity measurements which
            have been identified as insecure and thus rejected. 
            The reasons are written at the left-upper part of each panel.
            The symbols and lines are the same as in Fig.~\ref{f_rtv}.
   }
   \label{f_rtv_bad}
\end{figure}

In Fig.~\ref{f_rtv} we show the observed 
rotation curves of the 33 galaxies in the final
sample,  together with the fitted models,  in order to give a visual
impression of the quality of the data and the fits.
We note that a few galaxies may have irregular
kinamatics (e.g. C0016\_P1 11\_A). 
Examples of objects
rejected by our quality-control procedure are shown in Fig.~\ref{f_rtv_bad}.
The rejected objects were not used in our analysis.

The measured rotation velocities were corrected for
inclination multiplying them by $(\sin i)^{-1}$.
The uncertainties in the final values were estimated taking into account 
the errors in the
inclinations from GIM2D, and the errors 
in the rotation velocities from ELFIT2PY.
Table~\ref{t_nem} provides some statistics for the spectroscopic data.

The internal consistency of the rotation velocities was checked using four
galaxies observed in two different masks. Three of them showed emission lines.
After the quality-control process, two galaxies were left with only one
acceptable  spectrum due to slit angles being more than 60 degrees and/or
having rotation curves of insufficient quality from the other spectrum. 
Thus, for only one galaxy we measured two 
good-quality rotation velocities from 
two independent spectra (M1621\_P1 12\_A and M1621\_P2 10\_A). 
These rotation velocities were $163_{-8}^{+12}$ and
$186_{-14}^{+12}$ (km s$^{-1}$) respectively (before inclination
correction).
These measurements were thus consistent ($\sim1.3\sigma$ discrepancy), and 
the second value was chosen for the final sample due to marginally 
better $S/N$ in the spectum. 

The excellent seeing of the Subaru observations and the relatively high
spectral resolution of the FOCAS data allowed us to test the  effect of the
seeing and the instrumental resolution in the measurement of the rotation
velocities. This is important when comparing with the VLT results (cf. B05A),
obtained in poorer seeing and with slightly lower spectral resolution. 
The spectra of the surviving 33 galaxies
were convolved with Gaussians to 
simulate a $\sim 1.0$ arcsec seeing and a spectral resolution
$R \sim 950$. These values roughly correspond to the 
VLT data of B05A. The rotation velocities
were then measured in the same way. 
The ratio of the rotation velocity determined from the convolved
spectrum to the original one was $1.02\pm0.03$, and no systematic
trend was found as function of fitted emission scale-length.
Hence we should be able to compare our data with that of B05A
without any correction. We note that this applies only to the
galaxies with secure rotation velocities in our sample, and does not
necessarily mean that the method we have used to measure rotation
velocities is not affected by the seeing conditions.

Among the seven galaxies that we have in common with the VLT data
of B05A (Sec.~\ref{s_absmag}), three galaxies yielded 
secure rotation velocities. Unfortunately, only one of these, 
M2053\_P1 07\_A
in Fig.~\ref{f_rtv}, was measured securely by B05A. Our velocity estimate
for this galaxy was $182_{-17}^{+19}$ km s$^{-1}$ before inclination
correction, while B05A obtained $154_{-10}^{+11}$ km s$^{-1}$.
Although this is our only
direct comparison, the results are consistent with
each other within $1.4\sigma$.
The fully-corrected magnitudes, rotation velocities, and colours
of our sample are summarized in Table \ref{t_sample}.

\begin{table*}
 \centering
 \begin{minipage}{140mm}
  \caption{Summary of the emission line measurements.}
  \begin{tabular}{lrrrcrrcrr}
  \hline
          & \multicolumn{3}{c}{Observed}    &
          & \multicolumn{2}{c}{Emission-}   &
          & \multicolumn{2}{c}{V$_{rot}$-}     \\
          & \multicolumn{3}{c}{}            &
          & \multicolumn{2}{c}{detected}    &
          & \multicolumn{2}{c}{measured}       \\
  Cluster & N$_{c}$\footnote{Number of galaxies in clusters.}
          & N$_{f}$\footnote{Number of galaxies in field.}
          & N$_{u}$\footnote{Number of galaxies with unknown z.} &
          & N$_{c}$ & N$_{f}$                                    &
          & N$_{c}$ & N$_{f}$                                      \\
  \hline
  A2390           & 16 &  8 &  2  & &    8 &  6 & &  4 &  1 \\
  MS1621.5$+$2640 & 11 & 18 &  3  & &    5 & 17 & &  3 & 10 \\
  MS0015.9$+$1609 & 17 & 16 &  3  & &   11 & 15 & &  5 &  5 \\
  MS2053.7$-$0449 &  6 & 12 &  0  & &    4 & 11 & &  1 &  4 \\
  \hline
  Sub-total       & 50 & 54 &  8  & &   28 & 49 & & 13 & 20 \\
  Total           & \multicolumn{3}{c}{112}
                                  & &   \multicolumn{2}{c}{77}
                                                & & \multicolumn{2}{c}{33}\\
  \hline
\end{tabular}
\label{t_nem}
\end{minipage}
\end{table*}

\begin{table*}
 \centering
 \begin{minipage}{170mm}
  \caption{The data for the 104 galaxies in our sample 
           with measured redshift.
           The galaxies are sorted by redshift first for the
	   cluster members and then for  
	   the field galaxies.
           The columns are: (1) cluster, mask and slit IDs; (2) R.A. and
           (3) Dec.; (4) redshift; (5) cluster (C) or field (F) membership 
	   flag; (6)
           sample status (`TFR'$\equiv\,$in the final TFR; 
	   `em'$\equiv\,$emission-line galaxy without secure rotation
           velocity; `abs'$\equiv\,$absorption-line galaxy); 
	   (7) inclination (90$\degr\equiv$edge-on);
           (8) absolute rest-frame $B$-band magnitude, (9) rotation velocity,
           and (10) rest-frame $B-V$ colour.
  }
  \begin{tabular}{cccccccccc}
  \hline
Cluster/Mask/Slit & R.A. & Dec. & $z$ & Mem. & TFR & $i$ & $M_{B}$ & $\log V_{\rm rot}$ & $B-V$\\
                 &[J2000]&[J2000]&    &      &     &[deg]& [mag]   & [km s$^{-1}$]      & [mag]\\
  \hline
A2390 P2 00\_A & 21 53 45.6 & $+17$ 41 48 &  0.2147 &       C &    em. & 68 & $-21.99\pm0.06$ & ...                    &  0.59\\
A2390 P1 01\_A & 21 53 37.6 & $+17$ 44 10 &  0.2168 &       C &    TFR & 77 & $-20.90\pm0.11$ & $2.29_{-0.02}^{+0.01}$ &  0.81\\
A2390 P1 00\_A & 21 53 42.9 & $+17$ 40 07 &  0.2211 &       C &   abs. & 39 & $-22.07\pm0.05$ & ...                    &  0.57\\
A2390 P2 03\_A & 21 53 42.8 & $+17$ 41 53 &  0.2230 &       C &    em. & 71 & $-20.95\pm0.09$ & ...                    &  0.82\\
A2390 P1 08\_A & 21 53 31.6 & $+17$ 39 31 &  0.2230 &       C &   abs. & 69 & $-19.21\pm0.21$ & ...                    &  0.77\\
A2390 P2 10\_A & 21 53 28.9 & $+17$ 40 06 &  0.2240 &       C &   abs. & 73 & $-19.51\pm0.21$ & ...                    &  0.71\\
A2390 P1 03\_A & 21 53 34.9 & $+17$ 41 01 &  0.2280 &       C &   abs. & 58 & $-19.90\pm0.05$ & ...                    &  0.33\\
A2390 P2 05\_A & 21 53 47.0 & $+17$ 40 47 &  0.2301 &       C &   abs. & 78 & $-20.92\pm0.09$ & ...                    &  0.81\\
A2390 P2 09\_A & 21 53 30.0 & $+17$ 42 50 &  0.2304 &       C &   abs. & 82 & $-20.08\pm0.08$ & ...                    &  0.60\\
A2390 P1 06\_B & 21 53 35.2 & $+17$ 41 51 &  0.2318 &       C &   abs. & 55 & $-20.82\pm0.05$ & ...                    &  0.86\\
A2390 P1 09\_A & 21 53 33.4 & $+17$ 42 24 &  0.2322 &       C &    em. & 68 & $-19.91\pm0.05$ & ...                    &  0.35\\
A2390 P2 04\_A & 21 53 27.7 & $+17$ 43 13 &  0.2385 &       C &    em. & 35 & $-21.13\pm0.07$ & ...                    &  0.29\\
A2390 P1 10\_A & 21 53 38.1 & $+17$ 43 51 &  0.2401 &       C &    TFR & 78 & $-20.05\pm0.08$ & $2.28_{-0.01}^{+0.01}$ &   ...\\
A2390 P2 08\_A & 21 53 38.0 & $+17$ 43 47 &  0.2406 &       C &    TFR & 79 & $-20.33\pm0.08$ & $2.23_{-0.01}^{+0.01}$ &  0.66\\
A2390 P2 11\_A & 21 53 24.8 & $+17$ 42 56 &  0.2420 &       C &    TFR & 71 & $-20.07\pm0.07$ & $1.94_{-0.04}^{+0.06}$ &  0.50\\
A2390 P2 02\_A & 21 53 41.8 & $+17$ 42 15 &  0.2433 &       C &   abs. & 65 & $-21.04\pm0.10$ & ...                    &  0.81\\
M1621 P2 06\_A & 16 23 38.0 & $+26$ 35 39 &  0.4205 &       C &    TFR & 67 & $-20.75\pm0.08$ & $2.33_{-0.04}^{+0.03}$ &  0.56\\
M1621 P1 13\_A & 16 23 34.8 & $+26$ 30 45 &  0.4213 &       C &    TFR & 81 & $-20.11\pm0.08$ & $1.98_{-0.13}^{+0.06}$ &   ...\\
M1621 P1 00\_A & 16 23 38.9 & $+26$ 35 21 &  0.4216 &       C &   abs. & 69 & $-21.98\pm0.06$ & ...                    &  0.88\\
M1621 P2 04\_A & 16 23 39.1 & $+26$ 36 14 &  0.4217 &       C &    em. & 33 & $-20.93\pm0.09$ & ...                    &  0.40\\
M1621 P1 02\_A & 16 23 37.6 & $+26$ 33 01 &  0.4224 &       C &   abs. & 79 & $-21.48\pm0.07$ & ...                    &  0.93\\
M1621 P1 03\_A & 16 23 42.3 & $+26$ 30 55 &  0.4240 &       C &    TFR & 67 & $-21.15\pm0.07$ & $2.33_{-0.02}^{+0.01}$ &  0.79\\
M1621 P2 02\_A & 16 23 42.5 & $+26$ 33 44 &  0.4260 &       C &   abs. & 77 & $-21.68\pm0.07$ & ...                    &  0.96\\
M1621 P2 14\_A & 16 23 36.5 & $+26$ 35 01 &  0.4260 &       C &   abs. & 73 & $-20.98\pm0.08$ & ...                    &  0.84\\
M1621 P2 11\_A & 16 23 41.5 & $+26$ 35 36 &  0.4292 &       C &    em. & 65 & $-19.97\pm0.08$ & ...                    &   ...\\
M1621 P1 14\_A & 16 23 38.3 & $+26$ 34 28 &  0.4330 &       C &   abs. &  4 & $-21.06\pm0.06$ & ...                    &  0.90\\
M1621 P1 04\_A & 16 23 38.0 & $+26$ 36 10 &  0.4340 &       C &   abs. & 58 & $-20.97\pm0.07$ & ...                    &  0.65\\
C0016 P2 10\_A & 00 18 27.4 & $+16$ 23 03 &  0.5350 &       C &   abs. & 54 & $-20.55\pm0.10$ & ...                    &   ...\\
C0016 P2 12\_A & 00 18 40.1 & $+16$ 25 07 &  0.5377 &       C &    em. & 55 & $-20.58\pm0.12$ & ...                    &  0.77\\
C0016 P1 13\_A & 00 18 25.6 & $+16$ 24 39 &  0.5450 &       C &   abs. & 81 & $-20.95\pm0.11$ & ...                    &  0.96\\
C0016 P2 15\_A & 00 18 31.9 & $+16$ 24 41 &  0.5461 &       C &    em. & 79 & $-20.92\pm0.05$ & ...                    &  0.50\\
C0016 P2 01\_A & 00 18 38.9 & $+16$ 25 06 &  0.5490 &       C &   abs. & 66 & $-21.99\pm0.08$ & ...                    &  0.96\\
C0016 P2 00\_A & 00 18 30.9 & $+16$ 25 41 &  0.5492 &       C &    em. & 64 & $-22.28\pm0.08$ & ...                    &  0.59\\
C0016 P1 11\_A & 00 18 29.6 & $+16$ 26 36 &  0.5499 &       C &    TFR & 67 & $-20.63\pm0.16$ & $2.08_{-0.04}^{+0.03}$ &  0.40\\
C0016 P2 06\_C & 00 18 22.8 & $+16$ 26 05 &  0.5500 &       C &   abs. & 85 & $-21.35\pm0.09$ & ...                    &  0.82\\
C0016 P2 13\_A & 00 18 29.2 & $+16$ 23 11 &  0.5505 &       C &    TFR & 60 & $-20.44\pm0.11$ & $2.14_{-0.16}^{+0.12}$ &  0.55\\
C0016 P1 04\_A & 00 18 23.5 & $+16$ 25 09 &  0.5505 &       C &    em. & 57 & $-21.44\pm0.13$ & ...                    &  0.71\\
C0016 P1 05\_A & 00 18 31.1 & $+16$ 22 04 &  0.5506 &       C &    em. & 61 & $-21.34\pm0.10$ & ...                    &  0.69\\
C0016 P1 02\_A & 00 18 17.8 & $+16$ 23 23 &  0.5507 &       C &    TFR & 50 & $-21.65\pm0.18$ & $2.45_{-0.04}^{+0.04}$ &  0.82\\
C0016 P2 06\_B & 00 18 23.0 & $+16$ 26 05 &  0.5530 &       C &   abs. & 67 & $-19.23\pm0.07$ & ...                    &  0.70\\
C0016 P2 08\_A & 00 18 36.5 & $+16$ 25 15 &  0.5530 &       C &   abs. & 29 & $-20.60\pm0.10$ & ...                    &  0.40\\
C0016 P2 09\_A & 00 18 26.2 & $+16$ 25 08 &  0.5542 &       C &    em. & 47 & $-20.89\pm0.09$ & ...                    &  0.90\\
C0016 P2 14\_A & 00 18 35.6 & $+16$ 26 49 &  0.5572 &       C &    TFR & 54 & $-20.68\pm0.11$ & $2.24_{-0.06}^{+0.04}$ &  0.55\\
C0016 P2 16\_A & 00 18 34.0 & $+16$ 27 00 &  0.5600 &       C &    TFR & 81 & $-20.87\pm0.10$ & $2.33_{-0.05}^{+0.04}$ &  0.50\\
M2053 P1 06\_A & 20 56 20.2 & $-04$ 36 42 &  0.5793 &       C &    em. & 73 & $-21.24\pm0.05$ & ...                    &  0.73\\
M2053 P1 16\_A & 20 56 20.3 & $-04$ 38 21 &  0.5820 &       C &   abs. & 85 & $-21.20\pm0.21$ & ...                    &  0.71\\
M2053 P1 13\_A & 20 56 27.8 & $-04$ 34 48 &  0.5863 &       C &    em. & 55 & $-20.31\pm0.05$ & ...                    &  0.84\\
M2053 P1 14\_A & 20 56 22.1 & $-04$ 39 15 &  0.5863 &       C &    TFR & 84 & $-21.43\pm0.06$ & $2.24_{-0.04}^{+0.02}$ &  0.64\\
M2053 P1 11\_A & 20 56 18.5 & $-04$ 37 20 &  0.5880 &       C &   abs. & 50 & $-21.09\pm0.05$ & ...                    &  0.86\\
M2053 P1 04\_A & 20 56 19.5 & $-04$ 38 05 &  0.5909 &       C &    em. & 75 & $-22.47\pm0.21$ & ...                    &  0.67\\
  \hline
  \end{tabular}
\label{t_sample}
 \end{minipage}
\end{table*}

\begin{table*}
 \centering
 \begin{minipage}{170mm}
\centerline{Table~\thetable. Continued.}
  \begin{tabular}{cccccccccc}
  \hline
Cluster/Mask/Slit & R.A. & Dec. & $z$ & Mem. & TFR & $i$ & $M_{B}$ & $\log V_{\rm rot}$ & $B-V$\\
                 &[J2000]&[J2000]&    &      &     &[deg]& [mag]   & [km s$^{-1}$]      & [mag]\\
  \hline
M1621 P1 11\_A & 16 23 39.7 & $+26$ 31 07 &  0.0624 &       F &    TFR & 71 & $-19.06\pm0.07$ & $1.93_{-0.00}^{+0.00}$ &  0.46\\
M1621 P2 12\_A & 16 23 24.0 & $+26$ 34 14 &  0.0650 &       F &    TFR & 69 & $-19.16\pm0.05$ & $2.19_{-0.02}^{+0.01}$ &  0.71\\
C0016 P2 07\_A & 00 18 22.1 & $+16$ 23 45 &  0.0781 &       F &    TFR & 83 & $-16.64\pm0.11$ & $1.97_{-0.09}^{+0.08}$ &  0.42\\
M1621 P2 01\_A & 16 23 33.9 & $+26$ 30 55 &  0.1308 &       F &    TFR & 77 & $-18.94\pm0.06$ & $2.05_{-0.05}^{+0.05}$ &  0.67\\
C0016 P2 11\_A & 00 18 20.1 & $+16$ 26 00 &  0.1670 &       F &    TFR & 77 & $-17.86\pm0.07$ & $2.11_{-0.11}^{+0.08}$ &  0.46\\
M1621 P1 10\_A & 16 23 34.8 & $+26$ 35 45 &  0.1721 &       F &    em. & 71 & $-17.97\pm0.05$ & ...                    &  0.34\\
M2053 P1 02\_A & 20 56 22.3 & $-04$ 39 58 &  0.1854 &       F &    TFR & 73 & $-18.51\pm0.05$ & $1.80_{-0.03}^{+0.04}$ &  0.48\\
M1621 P2 00\_A & 16 23 46.2 & $+26$ 31 55 &  0.1910 &       F &    TFR & 55 & $-20.95\pm0.05$ & $2.26_{-0.03}^{+0.02}$ &  0.50\\
A2390 P1 11\_A & 21 53 33.9 & $+17$ 43 23 &  0.2454 &       F &    em. & 77 & $-21.13\pm0.10$ & ...                    &  0.43\\
A2390 P2 01\_A & 21 53 31.4 & $+17$ 41 34 &  0.2459 &       F &    em. & 63 & $-21.68\pm0.05$ & ...                    &  0.59\\
A2390 P1 02\_A & 21 53 38.2 & $+17$ 38 44 &  0.2500 &       F &    em. & 85 & $-20.38\pm0.08$ & ...                    &  0.34\\
C0016 P1 06\_A & 00 18 39.4 & $+16$ 24 18 &  0.2514 &       F &    TFR & 70 & $-19.26\pm0.11$ & $2.22_{-0.02}^{+0.01}$ &  0.55\\
M2053 P1 00\_A & 20 56 27.4 & $-04$ 35 27 &  0.2605 &       F &    TFR & 63 & $-19.95\pm0.05$ & $2.12_{-0.03}^{+0.03}$ &  0.50\\
C0016 P1 00\_A & 00 18 27.0 & $+16$ 26 58 &  0.2857 &       F &    em. & 58 & $-20.02\pm0.17$ & ...                    &  0.82\\
C0016 P1 12\_A & 00 18 31.4 & $+16$ 25 59 &  0.2862 &       F &    em. & 77 & $-19.71\pm0.11$ & ...                    &  0.53\\
C0016 P1 03\_A & 00 18 32.1 & $+16$ 25 21 &  0.3025 &       F &    em. &  0 & $-19.61\pm0.06$ & ...                    &  0.50\\
A2390 P2 06\_A & 21 53 32.8 & $+17$ 41 20 &  0.3390 &       F &    TFR & 73 & $-21.51\pm0.08$ & $2.44_{-0.04}^{+0.05}$ &  0.82\\
M2053 P1 05\_A & 20 56 20.2 & $-04$ 40 16 &  0.3447 &       F &    em. & 73 & $-20.17\pm0.05$ & ...                    &  0.78\\
M1621 P1 09\_A & 16 23 36.0 & $+26$ 34 17 &  0.3451 &       F &    em. & 82 & $-19.73\pm0.06$ & ...                    &  0.79\\
C0016 P2 19\_A & 00 18 18.2 & $+16$ 23 58 &  0.3501 &       F &    em. & 65 & $-20.11\pm0.09$ & ...                    &  0.43\\
M1621 P2 07\_A & 16 23 40.1 & $+26$ 34 04 &  0.3698 &       F &    TFR & 78 & $-20.15\pm0.09$ & $2.16_{-0.04}^{+0.05}$ &  0.73\\
M2053 P1 07\_A & 20 56 19.6 & $-04$ 38 47 &  0.3708 &       F &    TFR & 75 & $-20.56\pm0.05$ & $2.27_{-0.05}^{+0.04}$ &  0.65\\
C0016 P2 05\_A & 00 18 21.0 & $+16$ 26 13 &  0.3872 &       F &    TFR & 67 & $-20.22\pm0.07$ & $2.17_{-0.03}^{+0.03}$ &  0.37\\
C0016 P2 04\_A & 00 18 17.2 & $+16$ 25 34 &  0.3877 &       F &    em. & 43 & $-20.16\pm0.08$ & ...                    &  0.54\\
C0016 P1 14\_A & 00 18 15.8 & $+16$ 23 42 &  0.3878 &       F &    em. & 28 & $-19.83\pm0.11$ & ...                    &  0.27\\
M1621 P2 13\_A & 16 23 33.7 & $+26$ 35 45 &  0.3911 &       F &    TFR & 84 & $-19.74\pm0.06$ & $2.07_{-0.04}^{+0.04}$ &  0.46\\
A2390 P1 07\_A & 21 53 32.5 & $+17$ 42 48 &  0.3964 &       F &    em. & 63 & $-21.27\pm0.07$ & ...                    &  0.70\\
M2053 P1 09\_A & 20 56 32.5 & $-04$ 36 27 &  0.3973 &       F &    TFR & 58 & $-20.03\pm0.08$ & $2.29_{-0.03}^{+0.02}$ &   ...\\
C0016 P2 18\_A & 00 18 37.8 & $+16$ 24 55 &  0.3974 &       F &    em. & 14 & $-20.97\pm0.07$ & ...                    &  0.56\\
M2053 P1 01\_A & 20 56 14.7 & $-04$ 35 45 &  0.3983 &       F &    em. & 44 & $-21.08\pm0.05$ & ...                    &  0.40\\
M1621 P2 03\_A & 16 23 41.2 & $+26$ 32 02 &  0.4071 &       F &    em. & 76 & $-20.23\pm0.06$ & ...                    &  0.80\\
M1621 P1 06\_A & 16 23 39.0 & $+26$ 33 08 &  0.4130 &       F &   abs. & 79 & $-21.14\pm0.08$ & ...                    &  0.66\\
A2390 P2 07\_A & 21 53 44.7 & $+17$ 40 47 &  0.4160 &       F &   abs. & 69 & $-21.42\pm0.08$ & ...                    &  0.56\\
A2390 P1 04\_B & 21 53 41.9 & $+17$ 40 27 &  0.4360 &       F &   abs. & 52 & $-21.88\pm0.07$ & ...                    &  0.50\\
M1621 P2 08\_A & 16 23 44.2 & $+26$ 33 42 &  0.4378 &       F &    em. & 68 & $-20.35\pm0.09$ & ...                    &  0.52\\
M1621 P1 08\_B & 16 23 33.1 & $+26$ 33 40 &  0.4395 &       F &    em. & 62 & $-20.17\pm0.08$ & ...                    &   ...\\
M2053 P1 03\_A & 20 56 25.2 & $-04$ 35 59 &  0.4440 &       F &    em. & 37 & $-21.12\pm0.20$ & ...                    &  0.57\\
C0016 P2 02\_A & 00 18 15.6 & $+16$ 25 04 &  0.4471 &       F &    em. & 72 & $-21.26\pm0.07$ & ...                    &  0.41\\
C0016 P2 17\_A & 00 18 19.2 & $+16$ 25 43 &  0.4472 &       F &    TFR & 54 & $-20.14\pm0.08$ & $2.32_{-0.04}^{+0.05}$ &  0.37\\
M2053 P1 10\_A & 20 56 24.9 & $-04$ 37 37 &  0.4642 &       F &    em. & 42 & $-20.32\pm0.14$ & ...                    &  0.53\\
M1621 P1 01\_A & 16 23 42.6 & $+26$ 31 14 &  0.4762 &       F &    em. & 46 & $-21.51\pm0.07$ & ...                    &  0.56\\
M1621 P2 05\_A & 16 23 39.4 & $+26$ 30 58 &  0.4772 &       F &    TFR & 70 & $-21.25\pm0.08$ & $2.04_{-0.03}^{+0.02}$ &  0.38\\
A2390 P1 06\_A & 21 53 35.3 & $+17$ 41 55 &  0.5244 &       F &    em. & 35 & $-22.12\pm0.08$ & ...                    &  0.53\\
M2053 P1 08\_A & 20 56 18.7 & $-04$ 34 30 &  0.5289 &       F &    em. & 85 & $-21.61\pm0.05$ & ...                    &  0.83\\
C0016 P1 01\_A & 00 18 18.5 & $+16$ 24 01 &  0.5320 &       F &   abs. & 50 & $-21.24\pm0.11$ & ...                    &  0.96\\
M2053 P1 15\_A & 20 56 28.8 & $-04$ 39 40 &  0.5387 &       F &    em. & 51 & $-20.06\pm0.06$ & ...                    &  0.83\\
M1621 P1 05\_A & 16 23 47.5 & $+26$ 35 01 &  0.6300 &       F &    TFR & 44 & $-22.21\pm0.07$ & $2.43_{-0.10}^{+0.11}$ &  0.66\\
C0016 P1 09\_A & 00 18 32.8 & $+16$ 22 18 &  0.6550 &       F &    em. & 72 & $-21.79\pm0.08$ & ...                    &  0.82\\
C0016 P1 07\_A & 00 18 32.8 & $+16$ 26 09 &  0.6578 &       F &    em. & 74 & $-21.67\pm0.13$ & ...                    &  0.46\\
M1621 P2 09\_A & 16 23 31.9 & $+26$ 36 30 &  0.6608 &       F &    em. & 20 & $-21.34\pm0.08$ & ...                    &   ...\\
M1621 P2 03\_B & 16 23 41.2 & $+26$ 32 00 &  0.7006 &       F &    TFR & 57 & $-23.09\pm0.05$ & $2.58_{-0.04}^{+0.03}$ &  0.66\\
M1621 P2 10\_A & 16 23 32.8 & $+26$ 32 00 &  0.7407 &       F &    TFR & 56 & $-21.85\pm0.08$ & $2.35_{-0.04}^{+0.03}$ &   ...\\
M2053 P1 12\_A & 20 56 23.7 & $-04$ 37 04 &  1.0410 &       F &    em. & 72 & $-22.37\pm0.35$ & ...                    &  0.53\\
M2053 P1 13\_B & 20 56 27.8 & $-04$ 34 49 &  1.1950 &       F &   abs. &  9 & $-22.15\pm0.09$ & ...                    &   ...\\
  \hline
  \end{tabular}
 \end{minipage}
\end{table*}

\subsection{Morphologies}
\label{s_morph}

\begin{figure*}
\centering
\begin{minipage}{170mm}
   \includegraphics[width=150mm]{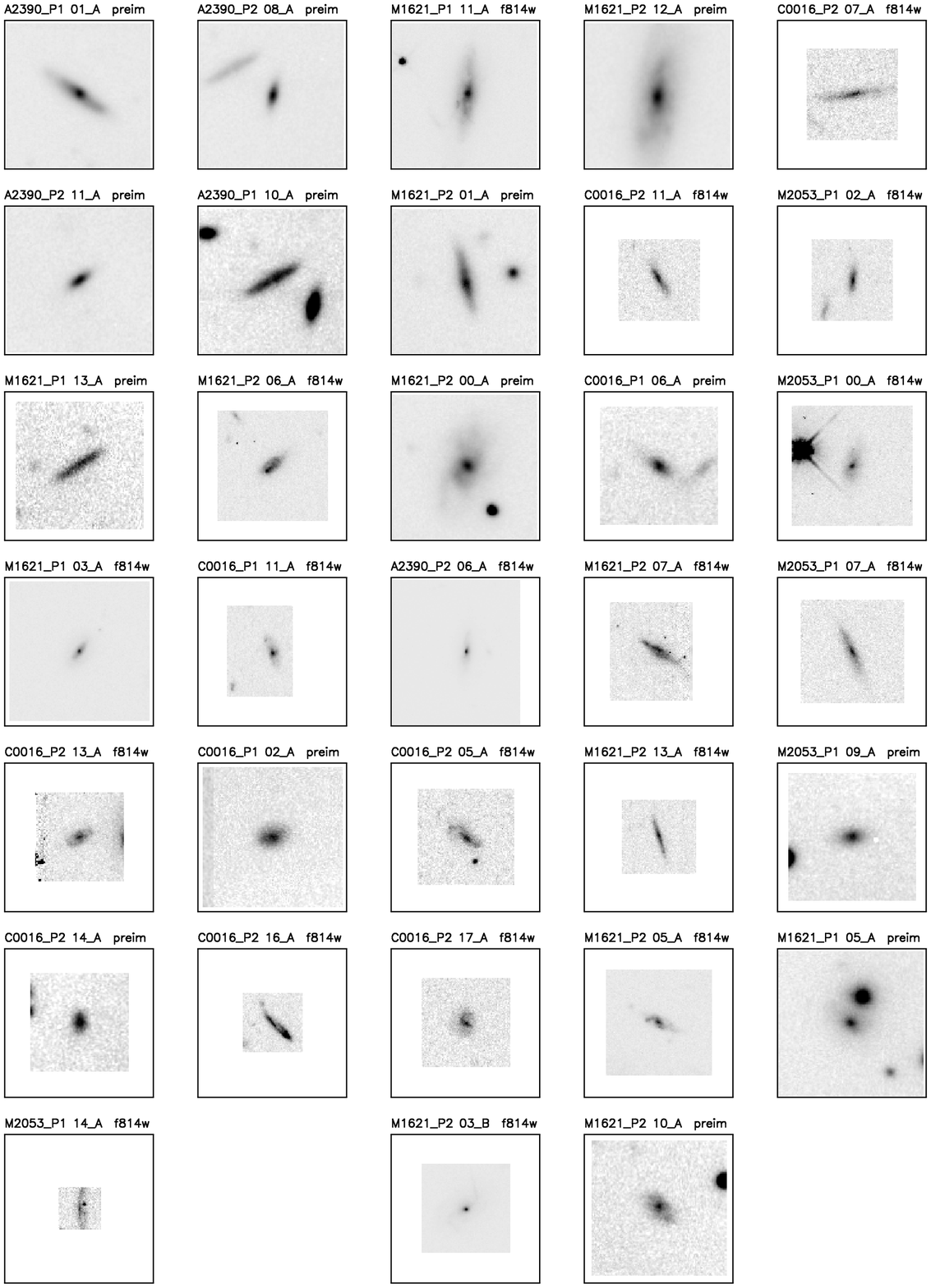}
   \caption{HST/FOCAS images of the galaxies in our final TFR sample. 
   We show HST images (labeled with band IDs) when available, 
   and FOCAS images (marked `preim') otherwise. The size of the box is
   $12^{\prime\prime}.2\times12^{\prime\prime}.2$, the same as the
   spatial dimension used in Fig.~\ref{f_rtv}.
   The panels are arranged as in  Fig.~\ref{f_rtv}.
   }
   \label{f_HST}
\end{minipage}   
\end{figure*}

The morphology of the galaxies was double-checked on the HST images after
the observations were completed. Although the FOCAS imaging allowed us to 
reliably select disk galaxies, it didn't have enough resolution 
to separate S0s from spirals. 
Among the 104 galaxies targeted, 62 had been imaged with HST.
Some statistics of our visual inspection are given in 
Table~\ref{t_morph}. As we targeted disk
galaxies (see Sec.~\ref{s_obs}), our sample includes no elliptical
galaxies except for one that happened to be on the slit of another 
target and was identified as a cluster member of A2390.
Sometimes it was difficult to distinguish S0s from spirals
even with the HST-images. Although we did not find any clear S0 galaxies, 
we did find some galaxies for which the classification was dubious. For that
reason, Table~\ref{t_morph} gives an upper limit to the number of S0s in our
sample. By design, most galaxies in our sample turned
out to be spirals. We estimate that at most $\sim$50\% of the galaxies with
absorption-line spectra were S0s. We will come back to this issue later. 
We note that all the galaxies in the final TFR sample (i.e., with secure 
rotation velocities) that had HST images had clear spiral morphologies,
as expected. Postage stamps of the HST/FOCAS images of the galaxies 
in the final sample are presented in Fig.~\ref{f_HST}. 

\begin{table}
   \caption{Visual Morphology of 62 galaxies for which HST images
            are available.}
   \begin{tabular}{lrrrrr}
   \hline
       & \multicolumn{2}{c}{Cluster}
       & \multicolumn{2}{c}{Field}     &       \\
       & $N_{\rm em}$ & $N_{\rm abs}$
       & $N_{\rm em}$ & $N_{\rm abs}$  & Total \\
   \hline
   Total &     15 &     13 &     34 & 0 &      62 \\
   S0    & $\la$1 & $\la$6 & $\la$2 & 0 &  $\la$9 \\
   \hline
   \end{tabular}
   \label{t_morph}
\end{table}

\section{Results}

The primary purpose of this study is to see whether spiral galaxies in 
clusters have higher star formation activity than those in the field. In the
following, we focus on three different aspects to investigate it: magnitude as
function of rotation velocity (i.e., TFR), colour, and fraction of spiral
galaxies with absorption-line spectra (i.e., no current star formation).

\subsection{Tully-Fisher relation}
\label{s_tfr}

\begin{figure}
   \includegraphics[width=84mm]{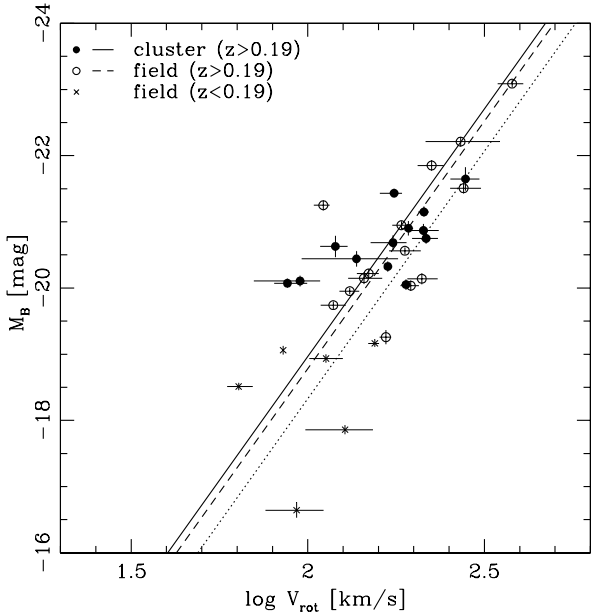}
   \caption{The TFR of the 33 galaxies for which 
            reliable rotation
            velocities were measured.
            Filled and open circles are the cluster and
            field galaxies at $z>0.19$. Field galaxies
            at $z<0.19$ (crosses) are kept separated because
            of possible sampling bias (see Sec.~\ref{s_absmag}).
            The dotted line shows the local TFR of PT92.
            The solid and the dashed lines show
            the linear fits to the cluster and the field galaxies
            at $z>0.19$ with the slope fixed to that of PT92.
   }
   \label{f_tfr}
\end{figure}

The TFR of our sample is shown in Fig.~\ref{f_tfr}.
There are 6 field galaxies at $z<0.19$. Although the
number is small, they are consistent with the local TFR
of  Pierce \& Tully (1992; PT92 hereafter; 
$M_{B} = -7.48 \log V_{rot} - 3.10 - 0.27$\footnote{
The intercept of the TFR in PT92 may be $0.45\pm0.12$ mag brighter
if the difference between $H_{0}$ that was derived from the
sample of PT92 (Pierce 1994, 86 km s$^{-1}$ Mpc$^{-1}$)
and the one used in this paper is taken into account
(B05B).
Our conclusions are not affected by this issue.}).
This suggests that our low redshift data is consistent with the results of
PT92. 
The galaxies at $z>0.19$ tend to be brighter than the local TFR of PT92 when
compared at the same rotation velocity. This suggests a possible luminosity
evolution at higher redshifts. More on this later. 

No clear difference  is apparent between the
cluster and the field galaxies at $z>0.19$. Using a simple  $\chi^{2}$
minimization method with internal errors taken into account 
(cf. B05B), we fitted linear regression  lines with the same slope as
the local TFR of PT92 to the field and cluster galaxies.  We find that 
the cluster
and the field galaxies are  $0.64\pm0.24$ and $0.46\pm0.23$ mag brighter,
respectively, than the local spirals in PT92.  The cluster galaxies appear to
be $0.18$ mag brighter than the field galaxies, but the difference has no
statistical significance.

\begin{figure}
   \includegraphics[width=84mm]{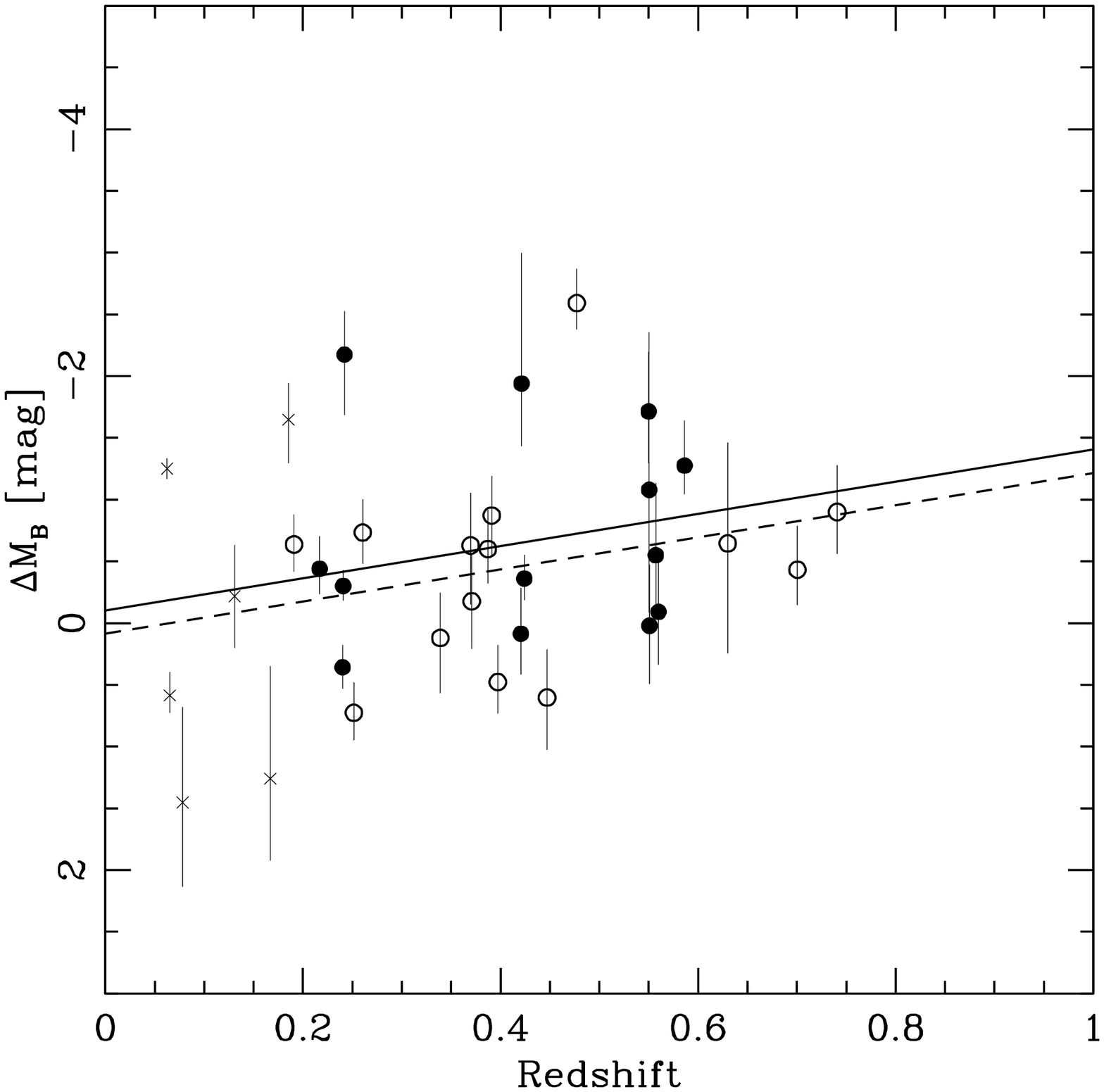}
   \caption{The TFR residuals from the local relation of PT92
            as function of redshift. The symbols are the same as in
            Fig.~\ref{f_tfr}. The dashed line shows a linear
            fit to the field galaxies at $z>0.19$. The solid line
            is the same but for the cluster galaxies with the
            slope fixed to that of the field galaxies.
   }
   \label{f_zev}
\end{figure}

An obvious problem with the above analysis is that our TFR has been derived
for galaxies at different redshifts.  If there is luminosity evolution with
redshift, it should be taken  into account. To assess this issue, 
in Fig.~\ref{f_zev} we plot the residuals (in magnitude) of 
our TFR from the local one of PT92 as function of redshift.
There is a trend suggesting positive luminosity
evolution with redshift. A linear
fit to the field galaxies at $z>0.19$ gives
\begin{equation}
\Delta M_{B} = ( -1.30 \pm 1.04 ) \ z + 0.09 \pm 0.46.
\label{e_zev}
\end{equation}
The median redshift of the sample used is $z=0.39$.
The fitted line has $\Delta M_{B}\sim0$
at $z=0$, consistent with the local TFR of PT92. The six field
galaxies at $z<0.19$ are also consistent with the fit.
Even taking into account
the possible luminosity evolution with redshift, Fig.~\ref{f_zev}
also suggests that there is no obvious difference in the 
behaviour of the cluster galaxies from that of the field galaxies.
The intrinsic standard deviations around the TFR linear fit,
after the luminosity evolution with redshift is
taken off, are 0.73 mag and 0.78 mag for the cluster and the field
galaxies, respectively. Hence, there is also little difference in the scatter
of the TFR of cluster and field galaxies. For our sample, this scatter
is about twice that of the local galaxies in PT92.
This is probably due to the fact that 
local TFR studies concentrate on samples of nice, regular spirals, 
while the high-redshift samples are, perforce, less clean.

\begin{figure}
   \includegraphics[width=65mm,angle=-90]{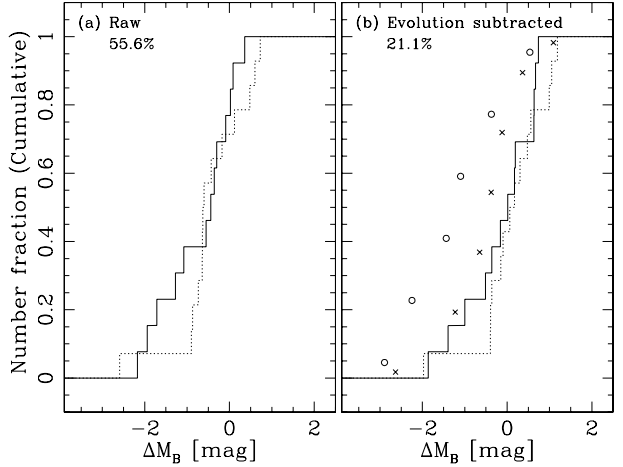}
   \caption{The cumulative distribution of the TFR residuals
            from the local relation of PT92 ($z>0.19$ galaxies only).
            The solid and the dashed lines in panel (a) show 
            the cluster and the field galaxies, respectively.
            The number at the left-top indicates the significance
            of the difference between the cluster and the field
            distributions from a Kolmogorov-Smirnov test.
            Panel (b) is the same as panel (a) but the 
	    measured luminosity 
            evolution with redshift is subtracted using
            Eq. (\ref{e_zev}). The right 
	    panel also shows the
            cluster (circles) and field (crosses) galaxies from the 
            'matched' sample of B05A after subtracting
            the evolution in the same way (see text for details). 
   }
   \label{f_dmh}
\end{figure}

In Fig.~\ref{f_dmh} we show the cumulative distribution
of our TFR residuals from the local relation of PT92.
We exclude the galaxies at $z<0.19$ to avoid possible
biases (although the 
results would not change if we included them).
The distribution of the cluster galaxies
resembles that of the field galaxies. The same conclusion 
is observed after the measured luminosity evolution with
redshift is taken off by Eq. (\ref{e_zev}). A Kolmogorov-Smirnov
test indicates that the significance of the difference
between the cluster and the field galaxies 
is 56\% for the raw values, reducing to 21\%
after the evolution is subtracted.

One might suspect that in the process of rotation velocity determination and 
quality-control  (Sec.~\ref{s_rtv}) we may have missed a number of cluster
galaxies that are undergoing a strong star-formation episode with irregular
kinematics. A simple test for this possible effect is to compare the fraction
of emission-line galaxies  which have secure rotation velocities for the cluster
and the field samples. According to
Table~\ref{t_nem}, the respective fractions for cluster and field galaxies
are $46\pm13\%$ and $41\pm9\%$, using Poissonian errors
(see also Ziegler et al. 2003).
Since these fractions are very similar, there is no
clear indication that such a selection bias is introduced by 
our quality control. 
In conclusion, there is no statistically-significant 
difference in the TFR between the cluster and field galaxies
of our sample.

\subsection{Rest-frame $B-V$ colours}

\begin{figure}
   \includegraphics[width=65mm,angle=-90]{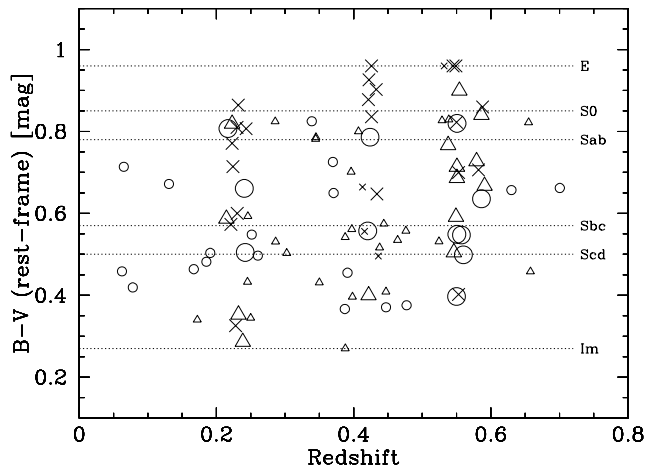}
   \caption{Rest-frame $B-V$ colour distribution of our sample
            galaxies. We plotted  93 of the 104 galaxies
            with known redshift, after dropping 9 galaxies with no
            colour information (see Sec.~\ref{s_absmag}) and two galaxies 
            at $z>0.8$.
            The symbols are the same as Fig.~\ref{f_z_B}.
            The dotted lines show the colours of E, S0, Sab,
            Sbc, Scd, and Im types at $z=0$ from Fukugita et al.
   }
   \label{f_col}
\end{figure}

We have seen that the TFR of the cluster galaxies in  our sample looks
similar to that of the field galaxies. In the previous section
we found no evidence for a selection bias against star-burst galaxies
introduced by our quality-control process.
A further test for possible selection biases can be carried out
if we look at the colour distributions of our samples, since
galaxies with a strong
star-formation episode would be very blue.
Fig.~\ref{f_col} shows the  rest-frame $B-V$
colour (Johnson-Morgan system) for all the galaxies in our sample with 
redshift and colour information available.
The rest-frame colours were derived as described in Sec.~\ref{s_absmag}.
The results below do not change if we use observed
colours which are independent of K-correction effects.
In the figure, the cluster line-emission galaxies with rotation
velocities unavailable (large triangles) do not show
bluer colour compared with those with rotation
velocities available (large circles). The mean rest-frame $B-V$ colour
of these galaxies are $0.63\pm0.05$ and $0.61\pm0.04$, respectively. 
Thus, the galaxies dropped in the rotation velocity measurement process are
not significantly bluer than the ones we kept. 
Hence it is unlikely that we have statistically dropped
cluster galaxies with higher star-formation from the rotation
velocity measurement.

The possible existence of such  
bias for the field galaxies used in our TFR can be
tested in the same way. The mean rest-frame $B-V$ colour
of the field galaxies in the TFR is $0.55\pm0.03$ mag, while
that of the field emission-line galaxies without secure
rotation velocities is $0.57\pm0.03$ mag. Thus the field galaxies
used in the TFR sample well the field emission-line galaxies 
in terms of their colour distribution.

\begin{figure}
   \includegraphics[width=84mm]{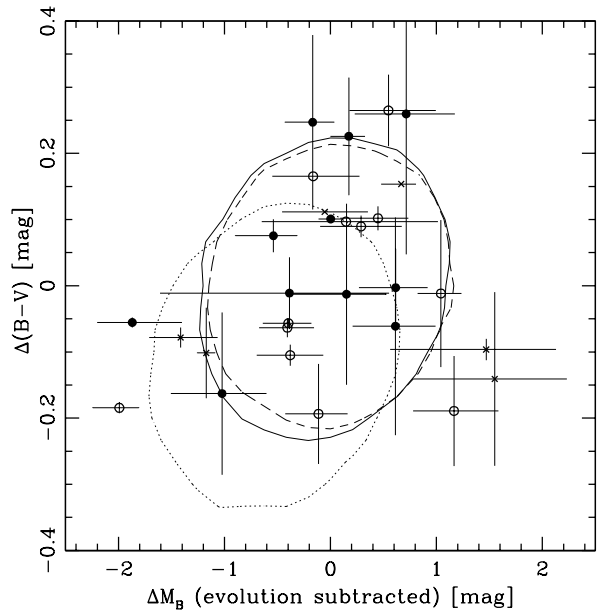}
   \caption{The TFR residuals vs. rest-frame $B-V$ colours
            for the 33 galaxies with secure rotation velocities.
            The redshift evolution  
            has been subtracted using Eq.~(\ref{e_zev}). 
	    The zero-point of the colours is set
            to the mean of the field galaxies. The symbols 
            are the same as Fig.~\ref{f_tfr}.
            The solid and dotted lines show the results
            of the simulation discussed 
	    in Sec.~\ref{s_sim}, indicating 
            the regions where 68\% of the cluster 
	    emission-line 
            galaxies are distributed in models 9 and 1
            respectively (see text for details). 
	    The dashed line is the same but
            for the field galaxies.
   }
   \label{f_simB_BV}
\end{figure}

The colours allow us a further consistency check of our TFR result
obtained in Sec.~\ref{s_tfr}. Higher star-formation rates 
would produce bluer colours at the same time as excess $B$-band 
luminosity. The average rest-frame $B-V$ colour of the cluster
galaxies used in the TFR is $0.61\pm0.04$, while for the field
galaxies it is $=0.55\pm0.03$. These values are only $1.2\sigma$
apart. If the cluster galaxies had higher star-formation than
the field ones (as suggested by B05A), 
we would expect them to be bluer, and not redder. 
Furthermore, Fig.~\ref{f_simB_BV} shows no correlation  between $\Delta M_{B}$
and $(B-V)_{\rm rest}$ for either the field or the cluster galaxies. Such a
correlation could be expected if excess star-formation were responsible for any
excess luminosity in these galaxies. The distribution of 
the cluster and field galaxies in this diagram is indistinguishable.  
These two points are consistent with the fact that the TFR of the
cluster galaxies in our sample is similar to that of the field
galaxies. 

As a final check,  if we include all the emission galaxies in our colour test
(i.e., not only those that make it into the TFR analysis) we still do not find
any significant difference  between the cluster and the field. The colour of
all the emission galaxies in the clusters ($0.62\pm0.03$) is slightly redder
than that of field galaxies ($0.56\pm0.02$), but the difference is again not
significant.  We will address quantitatively the colour and magnitude changes
expected in different evolutionary scenarios in Sec.~\ref{s_sim}.

\subsection{Spiral galaxies with absorption-line spectra}
\label{s_absorp}

The presence of emission lines is a direct diagnostic for current
star formation.   
The fraction of field spiral galaxies with no emission lines 
detected is $9\pm4\%$ (using Poissonian errors; see Table~\ref{t_nem}).
For all the cluster galaxies (including suspected S0s), 
the fraction is much higher, $44\pm9\%$. 
Even if we exclude {\it all\/} the suspected S0 galaxies 
and the serendipitous elliptical ($50\%$ of the cluster galaxies
with no emissions, Sec.~\ref{s_morph}),
we estimate that in our sample 
at least $29\pm9\%$ of the spiral galaxies in the
clusters have no emission lines.
Thus, there is an excess of non star-forming spiral galaxies 
in the clusters compared with the field. Our data suggest a factor
of at least $\sim3$ difference, but it could be as high as a factor of $\sim5$.
The existence of such 'passive' spiral galaxies in rich clusters
is consistent with the results of 
Dressler et al. (1999): if (at least some) 
spirals transform into S0s when they fall
into clusters, their star formation (and therefore their emission lines)
must be switched off at some point. 
This result, coupled with the lack of difference found  
in the TFRs of the cluster and field
spirals in our sample (Sec.~\ref{s_tfr}), seems to suggest that the 
process leading to the 
eventual cessation of star-formation is 
not accompanied by a drastic star-burst event. 
This point will be explored further in Sec.~\ref{s_sim}.

The majority of the 
cluster absorption-line galaxies have colours typical of early-type
galaxies, and the peak of their colour distribution is clearly different from
that of emission-line galaxies.  
The mean rest-frame $B-V$ colour of the absorption-line 
galaxies in our clusters is $0.76\pm0.04$, which is $0.20\pm0.05\,$mag
redder than the average colour of the field emission-line galaxies. 
A model star-burst containing 20\% of the total stellar mass 
and solar metallicity 
reaches $(B-V)_{\rm rest}\sim0.7$ about $1$ Gyr 
after the burst ended, and evolves very 
slowly thereafter  
($(B-V)_{\rm rest}\sim0.8$ at $\sim 2$ Gyr; Bruzual \& Charlot 2003).
A truncated star-formation model behaves similarly.
Thus the absorption-line galaxies in our sample have typically 
not experienced any significant star-formation for $\sim 1$ Gyr. 
However, there are a few absorption-line  galaxies
in our clusters that show bluer colours and are thus 
expected to be younger. This indicates that the sample 
of absoption-line disk galaxies in our clusters 
contains galaxies at different stages in their evolution.

If we limit the above analysis to those galaxies 
with bona-fide spiral morphologies as determined from 
the HST images, the mean $B-V$ colour of the absorption-line
galaxies in our clusters becomes $0.58\pm0.10$. This is closer to
the field emission-line galaxies, and would suggest a time
sequence for the morphological transformation from spirals to S0s:
spirals first switch off their star formation (i.e., no emission lines), and 
then become redder as their stellar populations age and their
morphologies are transformed.

\section{Comparison with previous work}
\label{s_comp}

\subsection{Detailed comparison}

\begin{figure}
   \includegraphics[width=84mm]{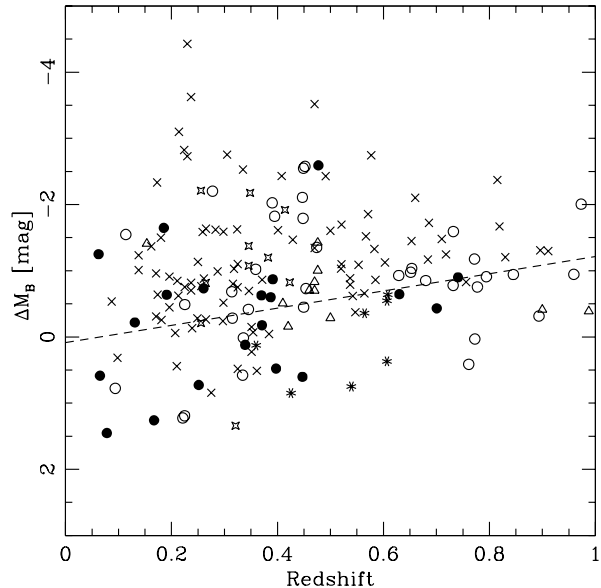}
   \caption{The TFR residuals 
            of our field galaxies vs.
            redshift compred with those obtained from other studies.
            The cosmology and the face-on $B$-band 
            extinction corrections are adjusted to match the ones adopted
            in this paper.
            The different symbols indicate: 
            field galaxies in our sample (filled circles);
            high-quality galaxies from Vogt et al. (1996, 1997) (triangles);
            Simard \& Pritchet (1998) (open stars);
            field galaxies from Ziegler et al. (2003) (asterisks);
            high-quality galaxies of B{\" o}hm et al. (2004) (open circles);
            and field galaxies from B05B (crosses). The dashed line
            is the same as in Fig. \ref{f_zev}.
            }
   \label{f_zev_cmp}
\end{figure}

Before we can consider the implications of our results it  is important to
assess  how reliable the rotation velocity measurements of intermediate
redshift galaxies are. To carry out an external check, we compare  our results
with those of other work which uses a similar technique at comparable
redshifts. We first compare the TFR of our field galaxies with the ones
derived  by other  workers in Fig. \ref{f_zev_cmp}. We plot the TFR residual
vs. redshift to separate redshift evolution from other effects.  We do not
compare our results with studies which use emission line widths (e.g., Rix et
al. 1997) instead of resolved rotation curves because that method is
substantially different from ours,  potentially causing significant systematic
differences.

Vogt et al. (1996, 1997) investigated the TFR of 16 field galaxies at $0.15\la
z \la 1$ selected based on $I$-band magnitudes from Forbes et al. (1996), the
Deep Extragalactic Evolutionary Probe project (Mould 1993), and the Hubble Deep
Field (Williams et al. 1996).
Eleven of the rotation curves obtained were labeled `high quality'
by them. 
The observed position-velocity diagrams were published 
for all the galaxies of their sample,
and our visual inspection criteria also class these 11 as good-quality
rotation curves. 
A visible turn-over point in the rotation curve was one of their
quality-control criteria, as it is for us. Their results are 
consistent with the ones we found for our field galaxies.

Simard \& Pritchet (1998) studied 12 field galaxies with
$0.26\le z \le0.43$. The sample was selected by its strong [O II]
emission (equivalent widths $>20$ {\AA}). The $B$-band
absolute magnitude range of their sample was similar to that of
Vogt et al. (1997).
The measured rotation velocities were not
corrected for inclination, so  
we applied an average correction  to their galaxies 
before placing them on Fig. \ref{f_zev_cmp} by assuming 
$\langle V_{\rm obs} \rangle=0.7854 \langle V_{\rm edge-on} \rangle$, 
which holds when galaxies are randomly oriented (cf. Simard \& Pritchet 1998). 
We also remove two
galaxies for which the derived error is larger than the derived rotation 
velocity. Since these authors did not publish
the observed position-velocity diagrams, we could not 
compare our visual quality-control with theirs. 
We do not know whether a 
visible turn-over point of the rotation curves was one of their
quality-control conditions.
The TFR residuals of their sample are, on average, slightly brighter than
in our field sample, although their redshift range
and their sample size are small.

Ziegler et al. (2003) investigated the TFR of~13 cluster and~7
field galaxies with $0.3\le z\le 0.6$ in three
cluster fields. All the observed position-velocity
diagrams were shown in J{\" a}ger et al. (2004), and 16 out of their 20
rotation curves pass our quality-control 
criteria, while four are rejected as rotation velocity overestimates.
These authors also required a visible turn-over point 
in the rotation curves as one of their acceptance criteria. 
The TFR residuals of their field galaxies are consistent with 
the local galaxies of PT92, even 
at intermediate redshifts. Therefore, their field galaxies
are slightly fainter than ours at a given rotation velocity, although
their sample size is small. 

B{\" o}hm et al. (2004)  investigated the TFR of 77 field galaxies with
redshifts between $0.1$ and $1$ in the FORS Deep Field.  This work includes the
galaxies  studied by  Ziegler et al. (2002).  The sample was selected in the
$R$ band, requiring that the spectral energy distribution of the galaxies was
later than E/S0. They selected 36 of their 
77 galaxies as high quality, and for 18 of them
the observed position-velocity diagram was shown. 
We agree with their quality assessment for these 18 galaxies,
which are probably their best.
A visible turn-over point of the rotation curve was one of their
conditions for high quality, while those galaxies with a smaller
extent or asymmetries were classified as low quality. 
The TFR residuals of their sample is consistent with that of
our field sample. Naturally, our Eq.~(\ref{e_zev}) agrees well with
their resulting fit [see their Eq.~(11)].

B05B,  using the same sample as B05A (which includes also the data published 
by Milvang-Jensen
et al. 2003), studied the properties of field spirals at 
intermediate redshift. Their sample contains 89 field galaxies with
$0.1 \la z \la 1$. The observed position-velocity diagrams of
six galaxies were shown as an example in B05A, and our 
quality-control accepts five of them, while one is rejected as an 
underestimate. The main 
difference in the quality-control process of B05A with ours is that they 
placed less weight on the condition of a visible turn-over point, because
ELFIT2PY should take that into account when estimating the errors
(see B05A for details). 
For reference, we rejected 57\% of the galaxies in 
our sample during the quality control process, 
while B05A rejected 47\% of theirs.
The TFR residuals from B05B are, at all redshifts, 
about 0.5 mag brighter than those from 
our sample, while the rate of redshift evolution (i.e., the slope in 
Eq.~\ref{e_zev}) is consistent.
On panel (b) of Fig.~\ref{f_dmh} we compare the distribution of the TFR
residuals from our Subaru sample with the 'matched' sample of B05A after
removing the redshift evolution in the same way. A Kolmogorov-Smirnov test
indicates that the distributions for field galaxies from both samples
disagree with 98\% confidence.

To finalise our comparison with previous work,  we consider TFR studies that
compare  cluster and field galaxies. Ziegler et al. (2003) found no difference
in the TFR between cluster and field galaxies, consistent with our result. We
note that the behaviour of both the field and cluster galaxies of Ziegler et
al. (2003)  in the TFR residuals vs. redshift diagram is rather similar to
that  of ours. On the other hand, B05A (including Milvang-Jensen 2003 data),
found that the average rest-frame $B$-band luminosity of cluster spirals was
significantly ($\sim 0.7$ mag) brighter than that of field ones at the same
rotation velocity. We do not find that effect in our Subaru data. Panel (b) of
Fig.~\ref{f_dmh} shows a comparison of our sample with that of B05A.

\subsection{More on quality control issues}

The visual quality control process is, perforce, subjective.  The main
difference found with some of the previous studies discussed above is the
treatment of the turn-over point of the rotation curve. We find  reasonable
consistency between  our results and those from studies that took the visible
presence of a turn-over point in the observed velocity-position diagram as a
requirement. It is therefore reasonable to expect that  differences between our
results and those from other studies  (e.g., B05A) could be due to this issue.

Requiring a visible turn-over point has both advantages and disadvantages from
a scientific point of view.  If we require it, we may drop from our sample  a
number of small galaxies which undergo strong star formation (cf. `downsizing'
scenario), or cluster galaxies which may have more concentrated star-formation
due to interactions with the cluster environment (cf. Moss \& Whittle 2000). 
On the other hand, if we estimate the rotation velocity of a galaxy using 
a region that is too small (without reaching the turn-over point), 
the dark matter may not yet dominate, thus underestimating it. 
(e.g., Verheijen 2001).
ELFIT2PY is designed to overcome this problem by direct
2-dimensional model spectrum fitting.
So far, however, the possible systematics 
introduced by the software when no visible turn-over point is visible 
have not been explored. This is the main reason why, in this paper,
we decided to use the visibility of the turn-over point as one of
our quality-control conditions.

\begin{figure}
   \includegraphics[width=84mm]{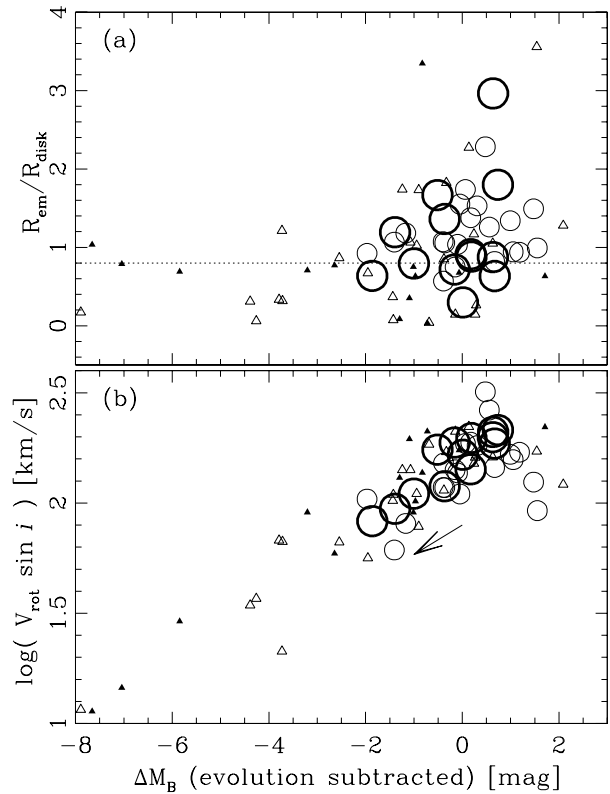}
   \caption{Panel (a): the TFR residuals as a function of the 
            the ratio between the fitted emission-line 
	    scale-length ($R_{\rm em}$) and the photometric 
	    disk scale-length ($R_{\rm disk}$) for 
	    all the galaxies with detected 
	    emission lines in the Subaru sample. 
	    The redshift evolution was subtracted 
	    using Eq.~(\ref{e_zev}).
            Thick large and thin small circles
            denote the cluster and field galaxies with secure rotation
            velocities, respectively. The solid and open triangles
            represent the cluster and field galaxies with 
	    unreliable (i.e., rejected) measurements.
            The dotted line indicates where $R_{\rm em} = 0.8\times 
	    R_{\rm disk}$.
            Below this line most measurements have been rejected (triangles).
            Panel (b): the TFR residuals as function of rotation
            velocity. The correction for inclination in the velocity
            is taken off to see the raw measured values.
            The arrow shows how an error in the 
            rotation velocity changes the position of
	    galaxies in this diagram.
            The direction and the dimension of the arrow corresponds
            to a 36\% underestimate in the rotation velocity at a fixed
            magnitude, which results in making the TFR residuals 1 mag
            brighter.
   }
   \label{f_r_rtv}
\end{figure}

Detailed model simulations to test the behaviour of ELFIT2PY under different 
conditions are being carried out, and will be discussed elsewhere (Bamford et
al. in preparation).  Nevertheless,  we have found that galaxies with small
fitted emission-line scale-length tend to have brighter TFR residuals.
Fig.~\ref{f_r_rtv}(a) shows the Subaru TFR residuals as a function of the ratio
between the fitted emission-line scale-length and the photometric
disk scale-length. We show both 
the galaxies with secure rotation velocity determinations
{\it and\/} those with insecure ones (i.e., the ones that did not pass   
our quality-control).
The photometric disk scale-length is derived using GIM2D on 
the redest HST images, when available,
and pre-imaging otherwise. 
If we look at all the emission-line galaxies in the
figure, there is a trend in the sense that  objects with smaller emission
scale-lengths have brighter offsets in the TFR residuals. 
A possible explanation for this trend is that ELFIT2PY may tend to
underestimate the rotation velocity for galaxies with smaller emission
scale-lengths, but it could well be that these differences are real.   On the
other hand, the galaxies with {\it secure\/} rotation velocities (i.e., the
ones used in the analysis)  do not show such a trend, so this gives us some
confidence in our quality-control/rejection process.  Nevertheless, if
concentrated emission is linked with increased star star formation (and thus
brightening),  eliminating galaxies with compact emission from our analysis
could  hide some interesting and important physical effects.  It is clear that
using the presence or absence of a  turn-over point in the rotation curve as a
quality-control  criterion is not as clear cut as one could naively expect, and
it could significantly affect the conclusions.

In our sample, a significant fraction of the galaxies with a fitted emission
scale-length smaller than the photometric disk scale-length tend to have
insecure rotation velocity measurements. Many of them were rejected  because
they clearly underestimated the rotation velocity and/or have strong nuclear
emission (cf. \S\ref{s_rtv}).
It could be argued that an acceptance threshold should be introduced in
the emission-line scale-length.
A reasonable
threshold could be set by 0.8$\times$$\langle$photometric 
disk scale-length$\rangle$
if our quality-control can be taken as optimal (cf. Fig.~\ref{f_r_rtv}a).
However, we decided not to impose such a strict limit and to 
use visual inspection to reject/accept measurements because we may lose some
galaxies whose rotation velocities can be determined reasonably well
but which have more centrally-concentrated star formation. These galaxies 
could provide important clues of the evolutionary processes
taking place in the cluster environment (Moss \& Whittle 2000; Milvang-Jensen
2003). This issue will be discussed in a future paper. 

Fig.~\ref{f_r_rtv}(b) shows an apparent correlation between
the TFR residuals and the rotation velocity.
We note that this correlation would be the natural
consequence of the TFR residuals being a function of
the rotation velocity [$\Delta M_{B} = 7.48 \log V_{rot} + \dots$].
The arrow on the panel shows the direction of the correlated errors.
Although B{\" o}hm et al. claimed the same trend at
$\log V_{rot} > 1.8$ as evidence of
mass-dependent evolution (using a bootstrap bisector fit),
the trend can also be explained by random errors in the
rotation velocities for a magnitude-limited sample without requiring any
evolution (see B05B).
The insecure rotation velocity measurements tend to occupy the lower
velocity region of this trend. This may be an indication that the
rotation velocities of these galaxies could be systematically underestimated,
although is is not possible to rule out that a real effect is present.

To summarise the above discussion, the subjectivity of the quality-control
process,  especially the treatment of the turn-over point of the rotation
curve, can significantly change the TFR results, and it could be the source of
discrepancies  between different studies.  It is thus important to present the
data,  including the observed position-velocity diagrams, in such a way that 
cross-checks between different samples and studies can be carried out.  Of
course, we must not forget other possible sources of discrepant  results such
as different instrument/telescope combinations or  real physical differences
between samples.

The discussion here illustrates the difficulties in comparing datasets
observed by different groups and/or using different instruments. Even
with the utmost care, inconsistencies are often unavoidable.  But even
if we cannot compare the galaxies in our Subaru sample with the ones
in the VLT sample of B05A,  the differences between field and cluster
galaxies within each dataset should be more reliable because
their field and cluster samples were selected, observed
and analysed in exactly the same way.
Since we cannot rule out that the differences  
between cluster and field galaxies found by B05A may be real,
at this stage we will take their results at face value, as a working
hypothesis, and explore the implications. 

\section{Implications of our results}

We have seen no obvious difference between the cluster and the
field emission galaxies in the TFR. This result is consistent
with the results of Ziegler et al. (2003). 
We have also found no difference in the rest-frame $B-V$ colour
distributions of the field and cluster emission-line spirals.
On the other hand, we have detected a
difference in the fraction of absorption-line spirals in our field 
and cluster samples.
This is rather a puzzling situation, since this last
result is usually interpreted as evidence for a transformation
of spiral galaxies into S0s in cluster environments, 
while the first two results seem to suggest no difference in the
cluster and field emission-line spirals. 
It is not impossible that, for a spiral galaxy
sample like the one we have studied, the statistical effect  on the
colours and luminosities produced by the galaxy transformations we
have postulated is too weak to be detected.
This issue can only be addressed with detailed modeling.

The situation seems even more complicated if we accept the 
results of the study published by B05A, who found that cluster spirals 
are brighter than the field ones at the same rotation velocity,
suggestive of a period of enhanced star-formation
before it is switched off.  Why don't we detect the same phenomenon?
One possibility is that cluster-to-cluster differences
could play an important role, although both studies concentrate on relatively
rich clusters. Another possibility is that, due to the small number statistics,
our Subaru sample may have missed several cluster galaxies that are
undergoing this putative starburst, preventing us from detecting the same
effect found by B05A. The likelihood of such possibility
needs to be investigated. 

To explore these two issues, we have computed several simulations of the
expected changes in the statistical properties of field and cluster galaxy
samples under different plausible evolutionary scenarios.

\subsection{Model expectations}
\label{s_sim}

The basic assumption of our models is that field spiral galaxies fall into 
clusters and get transformed into S0s after their star formation is 
extinguished by the cluster environment (e.g., after their gas is removed by
ram-pressure stripping). We start by assuming that the parent field galaxy is
forming stars at a relatively constant rate.  The star-formation history of
such galaxy after it enters the  cluster will depend on the details of its
interaction with the  intra-cluster medium, the cluster tidal field, other
galaxies, etc.  One possibility is a complete and sudden truncation of the star
formation, as expected if {\it all\/} the gas is removed from the disk and halo
of the galaxy. A second possibility is that the star formation ceases gradually
(e.g., exponential decay), as expected if only the halo gas is removed while
the gas disk is retained. In this case,  the star formation will decay as the
available gas is used up and not replenished  from the halo reservoir. A third
possibility is that in the interaction process, the halo gas is removed, and
the disk gas is compressed, producing a period  of enhanced star-formation
leading to a rapid exhaustion of the gas reservoir.  Our models explore these
possibilities, and predict the observational consequences we would expect in a
study like ours.

\begin{figure}
   \includegraphics[width=65mm,angle=-90]{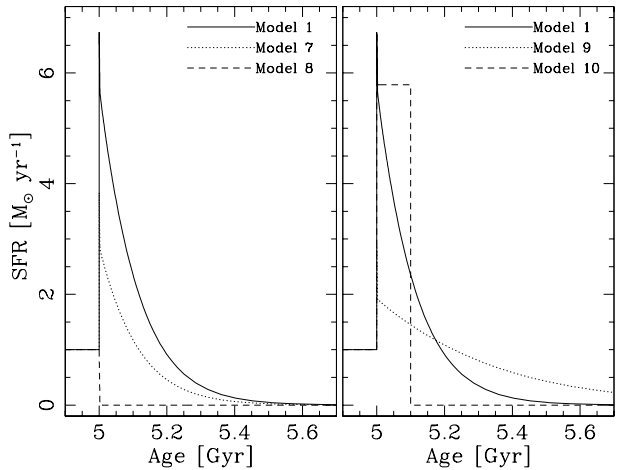}
   \caption{ Examples of star-formation histories explored in our  
             simulations. Model IDs correspond to
             those in Table~\ref{t_sim}. The absolute scale
             of the vertical axis is arbitrary.
   }
   \label{f_simsfr}
\end{figure}

For the field galaxies we assume stellar populations with constant star
formation and a fixed age of 5$\,$Gyr. The results are not sensitive to this
value provided that is larger than a few Gyrs. For the cluster galaxies we use
the same model until the galaxy encounters the cluster.  At that time,
different star formation histories are considered, with different star-burst
contributions and timescales. Fig.~\ref{f_simsfr} shows some examples of the
star-formation histories we used.   The star formation after the
encounter with the cluster has an $e$-folding timescale $\tau$, with burst mass
fraction $f_{b}$ relative to the underlying population.   
Our calculations assume  solar metallicity  for the stellar populations, and
use  Bruzual \& Charlot's (2003) models with a Chabrier (2003) initial mass
function and the Padova 1994 stellar evolutionary tracks. 
Table~\ref{t_sim} lists the values  of the adopted model parameters (see below).

We further assume that the parent field galaxies 
have a rest-frame $B$-band luminosity function like that determined from
the 2dF survey 
by Norberg et al. (2002) for the local universe
($\alpha=-1.21$ and $M^{*}_{B}=-20.43$ with $h=0.7$),
and make $M^{*}_{B}$ evolve with redshift according to Eq.(\ref{e_zev}), while
keeping $\alpha$ constant. 
We select galaxies randomly from this luminosity function
to generate a field population and a cluster population whose star-formation
histories
come from the models described above. We assume 
magnitude uncertainties of 0.1 mag, similar to the ones in our data. 
The magnitude limit of the model samples is 
set by the observed $R$-band magnitudes to mimic our observational selection.
At each redshift, this $R$-band limit is translated into a value of
$M_{B}(z)$ using our adopted cosmology and assuming the spectral energy
distribution of the field galaxy models. 
Because of the way our observed samples
were selected, this limit is reasonably close to $M^{*}_{B}$ for the 
`quiescent' field population. Obviously, we take 
into account that the star-formation history 
of a galaxy alters its luminosity and colours, 
bringing in or out of our simulated sample
depending on its apparent magnitude. 

We also assume that the
field spirals retain their spiral morphology and emission line
spectrum during the whole period covered by our simulations (i.e., they 
continue forming stars at a constant rate). 
This assumption seems reasonable for our sample 
since 91\% of our field galaxies
have detected emission lines (cf. Sec.~\ref{s_absorp}). In the case of the 
cluster galaxies, we assume that the spiral morphology
is observable for a time $\Delta t_{sp}$ after their entrance in the 
cluster/start of the possible burst. After that time, the cluster galaxies
do not enter our sample since they would not be classified as spirals anymore. 
That defines the maximum age of the 
cluster galaxies in our simulated samples.
It is important to remember that 
the absorption-line galaxies in our cluster sample could be
contaminated by S0s when considering the results of our simulations. 

Thus, a population of  
infalling cluster spirals is built in our simulations 
by randomly-sampling the field
luminosity function and assigning to the galaxies  an age $t_{b}$ counted from 
their entrance in the cluster/start of the burst. This age is taken randomly
from the interval ($0 \le t_{b} \le \Delta t_{sp}$). The  corresponding
magnitude change  in each band is determined from  the model. Obviously, if 
$t_{b}$ is comparable to or smaller than $\tau$, the galaxy would be brighter
than the parent galaxy because of the starburst. At later times, it would
become fainter due  to the cessation of star formation and the aging of the
stellar population. The magnitude limit determines whether the galaxy  enters
the sample or not.  If the galaxy ends up in the sample, a rotation velocity is
assigned to it using  the TFR of PT92 assuming luminosity evolution from
Eq.(\ref{e_zev}). In this process the luminosity that we use to  determine the
galaxy's rotation velocity is the one it had {\it before\/} it entered the
cluster. The underlying assumption is that the interaction with the cluster
environment alters the galaxies' luminosities (via a change of their star
formation histories), but not their masses/rotation velocities.  The error in
$\log V_{rot}$ (km s$^{-1}$) is set to 0.1, comparable with our data. A
comparison sample of field spiral  galaxies is built by randomly populating the
field luminosity function down to our apparent magnitude limits, and assigning
rotation velocities to them in a similar way.

We further define the parameter  $\Delta t_{em}$ ($0 \le \Delta t_{em} \le
\Delta t_{sp}$) by requiring that  emission lines are observables  only  if
$t_{b} \le \Delta t_{em}$. This parameter controls the fraction of
absorption-line  spiral galaxies. Note that this is not a free parameter since the
equivalent width (EW) of the emission lines (and therefore their detectability) 
depends on the relative intensity of the star formation at any given time,  and
therefore  it is possible to estimate reasonable values of $\Delta t_{em}$ for 
each model (see below).

The observed redshift $z_{obs}$ is set to 0.4, which is close to the median of
our sample (Sec.~\ref{s_z}). We set $\Delta t_{sp}=3$ Gyr, based on the
numerical simulations of gas stripping by Bekki, Couch, \& Shioya (2002), and
$\tau=0.1$ Gyr following Quilis et al. (2000).  For each model, $\Delta t_{em}$
is determined  by considering reasonable observational limits in the EW
of the emission lines.  The number of ionising (Lyman continuum)
photons, $N_{Ly}$,  is taken from the population synthesis models, and used  to
calculate the  H$\alpha$ luminosity as  $L(H\alpha)=1.36 \times 10^{-12} \cdot
N_{Ly}$ (erg s$^{-1}$). Our  calculations assume that all the ionising photons
from the  hot stars  are absorbed (case B recombination), an   electron
temperature $T_{e}=10^{4}\,$K and an electron density $n_{e}=100$ cm$^{-3}$.  
We further apply a mean extinction of $A_{V}=1.1\,$ mag to the emission lines, 
as  derived for
nearby SDSS spirals by  Nakamura et al. (2004). Seaton (1979) extinction law is
then used  to derive the EW of the  line from the line and continuum
luminosities of the models. The observational threshold is set to 10{\AA} for
H$\alpha$, which is more or less similar to that of our observations. Note that
because the extremely rapid  evolution of the EW after the star-formation has
ceased (see, e.g., Alonso-Herrero et al.\ 1996), a factor two difference in
the  EW limit does not affect $\Delta t_{em}$ significantly. Equally, setting
the limit using a different line would make very little different.  For the 
burst strength $f_{b}$ we use a reasonably high value of $0.2$ initially.    
The model calculated with  this set of parameters is called our `standard'
model (number 1). We will explore later the effect of varying these
parameters (cf.  Table~\ref{t_sim}), and the 
star formation histories (see Fig.~\ref{f_simsfr}).

We simulate cluster and field galaxy samples containing 1000 galaxies each. The
results of the simulations in terms of  observable quantities are summarized in
Table~\ref{t_sim}.  The models achieving relatively close values to the
observed ones are marked with stars. The values of the
scatter shown in the table represent the standard deviations
for the simulated galaxy 
samples\footnote{The error in the mean values can thus 
be obtained by dividing the
standard deviation by $(N-1)^{1/2}$, where $N=1000$
is the number of galaxies in the simulations.}, and the observational errors 
for the observed quantities. 
The standard deviations of
the predicted quantities for the simulated 
field galaxies depend only on the
assumed errors. As the table caption indicates, 
the standard deviation 
of $\Delta M_{B}$ for the simulated field galaxies
at $z=0.4$ is $0.77$, remarkably close to that of our 
observed field TFR ($0.78$; see Sec.~\ref{s_tfr}). 
For the cluster
galaxies, the scatter also depends on both the brightening of  galaxies due to
the burst  of star formation and the fading of galaxies in the fade-out phase
after the event. Nevertheless,  assuming errors comparable to those of our
observations, we find that the scatter in the observed properties of  field and
cluster galaxies are almost the same, and consistent with the observed ones.
Since we are interested in relative differences between the cluster and the
field galaxies,  we  ignore possible offsets between the  absolute colours of
the models and the observed galaxies.  For simplicity, we also ignore the fact
that actual spiral galaxies show a wider spread in colour than our models due
to  differences in morphology, inclination, extinction, stellar populations,
metallicities, etc. We expect differential quantities to be more robust   
than absolute ones.  

As a first step, we explored the effects of varying some of the  parameters
on the predicted differences between the cluster and field model galaxies.
We found that the dependence of the model results on the actual value of
$z_{obs}$ is small (see Models 2 and 3).  This is due to the fact that our
observations were
designed to reach approximately the same absolute magnitude at each redshift.
Thus using our fiducial value $z_{obs}=0.4$ to model our complete dataset
should be safe. The value of $\Delta t_{sp}$ that best corresponds to our
Subaru sample is not well defined by our galaxy  selection procedure.
Luckily, changing this parameter  within very broad limits has almost
negligible effect on the predicted
differences between field and cluster model galaxies (Models 4 and 5).
Changing the magnitude limit of the observations has only a moderate effect
(Model 6): going deeper in the luminosity function
changes the results a little by including a few more galaxies in the 
fade-out phase in the galaxy samples. 

The main conclusion of exploring models 1--6 is that with the parameters
explored there we do not get results  which simultaneously agree with our
observed changes in luminosity, colour and absorption-line galaxy fraction. A
better match to our observations  is achieved  if we reduce the burst strength
$f_{b}$ to 0.1 (Model 7). A truncated star formation history without  a
star-burst (Model 8) also predicts close values to our observations in terms of
magnitude and colour differences  because cluster emission-line  galaxies spend
a very short time in the emission  phase, and thus do not have time to  make
their properties different from the field galaxies.  However, the fraction of
absorption-line galaxies ends up being too high for this model since the vast
majority of the cluster galaxies  would have stopped forming stars.
Alternatively, our observed values are roughly achieved also if the burst
timescale is as long as $\sim0.3$ Gyr (Model 9). 

The magnitude and colour distributions from Models~1 and~9 are over-plotted in
Fig.~\ref{f_simB_BV}. It seems as if Model~9 not only provides a better match 
to the average observed values, but also to their distributions. However, our
models tell us that  there is a degeneracy between the burst mass fraction and
the star-formation time scale. Models~7 and~9 reproduce the observations almost
equally well. Nevertheless one interesting conclusion is that it is possible to
have substantial starbursts in the cluster spirals without significantly
changing the average colours and luminosities of the population of
line-emitting spiral galaxies in a sample like ours.  It is also clear that the
scenario we propose would imply a significant increase in the fraction of
spiral galaxies with absorption-line spectra when the field galaxies fall into
the clusters, as observed.

On the other hand, B05A found that the mean TFR residuals
of the cluster galaxies are 0.7 mag brighter than that
of the field galaxies, 
in agreement with the earlier findings of 
Milvang-Jensen et al. (2003). This is closer to the predictions 
of more massive
or shorter timescale bursts (e.g., Model~1)
than our case. If we assume that the results
from B05A can be represented by Model~1, the difference from the Subaru
observations would be $\sim 1.1 \sigma$ in
terms of $\Delta^{c,em}_{f} ( \Delta M_{B} )$
($\sim 1.6 \sigma$ when compared with B05A directly as opposed to comparing 
with the model). However, the difference 
would be $\sim 4.5 \sigma$ in terms of 
$\Delta^{c,em}_{f} (B-V)$, i.e., the model predicts
significantly bluer cluster emission-line galaxies. 
Taken at face value, this implies a very low probability that the 
discrepancies in the results from the VLT and Subaru data 
are due simply to a different sampling of the same parent population. 
Of course, this assumes that one can apply simple Gaussian statistics to 
this complex problem, which is far from clear. 

Thus, we cannot rule out that other factors not accounted for in our
simulation could also play some role in the observed differences. The lower  
median redshift of the Subaru sample ($0.39$), as compared with the VLT one
($0.52$), would suggest that one would expect a stronger evolutionary effect
for the latter. However, B05A show that the effect is present for even their 
lower redshift clusters ($z\sim 0.3$).
Cluster-to-cluster differences are, of course, another
potential candidate. 

The fact that we get a relatively weak 
statistical significance for the inconsistency between 
the Subaru results and Model 1, or between the Subaru data 
and the VLT data of 
B05A in terms of
$\Delta^{c,em}_{f} ( \Delta M_{B} )$
is due to the large uncertainties associated 
with the measured values. Indeed, according to our
model simulations, one would need to determine
reliable rotation velocities for $\sim 50$ cluster spirals 
and  a similar number of
field ones to distinguish
Model 1 (strong, short burst) from Model 9 (weak burst with slowly-declining
star formation) at $3 \sigma$ level by using
$\Delta^{c,em}_{f} ( \Delta M_{B} )$ alone. Thus much larger samples than 
those presented here and in B05A are needed.

\begin{table*}
  \begin{minipage}{140mm}
   \caption{{\bf Model predictions for different parameters:} 
   The values of the model parameters for the standard model are
   discussed in the text. Only the parameters that change in each model
   are listed. The remaining parameters are as in the standard model.
   The models achieving values close to the observed
   ones are marked with stars.
   }
   \begin{tabular}{lcrrrrr}
   \hline
         & $\Delta t_{em}$
         & $\Delta^{c,em}_{f} ( \Delta M_{B} )$\footnote{Difference in TFR residuals between cluster and field
                                                   galaxies, 
                                                   and the {\it standard deviation\/} for the {\it cluster} 
						   galaxies (see text).
                                                   The average residual 
						   for the field galaxies and its standard deviation at $z=0.4$
                                                   is $\Delta M_{B,f}=-0.50\pm0.77$.
						   }
         & $\Delta^{c,em}_{f} (B-V)$\footnote{     Difference in rest-frame 
                                                   $B-V$ colour between the emission-line galaxies in the clusters
						   and the field, and the standard deviation for the 
                                                   {\it cluster\/} galaxies.
                                                   The average colour of the field galaxies and its standard deviation is
                                                   $(B-V)_{f}=0.34\pm0.14$.
						   }
         & $\Delta^{c,abs}_{f} (B-V)$\footnote{    The same as $b$, but for the absorption-line galaxies.}
         & $f_{c,abs}$\footnote{                   Fraction of absorption-line cluster galaxies. 
	                                           Note that the observed fraction shows the measured fraction 
						   of cluster absorption-line spirals minus
						   the measured fraction of field absorption-line spirals. 
						   This is done because our models assume that all the 
						   field spirals have emission lines.  } \\
   Model \# and Item                               & (Gyr) & (mag)          & (mag)          & (mag)          & (\%) \\
   \hline
   (1) Standard model                              & 0.26  & $-0.54\pm0.82$ & $-0.12\pm0.15$ & $ 0.16\pm0.19$ & $45$ \\
       \hspace{0.4cm} $z_{obs}=0.4$, $\Delta t_{sp}=3$ Gyr,    &  &         &                &                &      \\
       \hspace{0.4cm} $M_{\rm lim}=M^{*}_{B}(z)$, $f_{b}=0.2$, &  &         &                &                &      \\
       \hspace{0.4cm} $\tau=0.1$ Gyr                           &  &         &                &                &      \\
   \hline
   (2) $z_{obs}\rightarrow 0.2$                    & 0.26  & $-0.52\pm0.79$ & $-0.10\pm0.15$ & $ 0.26\pm0.22$ & $53$ \\
   (3) $z_{obs}\rightarrow 0.6$\footnote{ The nominal ages of the model for the cluster galaxies can
                                          exceed the age of the universe, although such galaxies
                                          are in the fade-out phase and thus insensitive to the
                                          assumed age of the underlying population at the time of the burst.}
                                                   & 0.26  & $-0.52\pm0.81$ & $-0.10\pm0.15$ & $ 0.15\pm0.20$ & $31$ \\
   \hline
   (4) $\Delta t_{sp} \rightarrow 1$ Gyr           & 0.26  & $-0.49\pm0.75$ & $-0.11\pm0.14$ & $ 0.14\pm0.17$ & $42$ \\
   (5) $\Delta t_{sp} \rightarrow 8$ Gyr$^{e}$     & 0.26  & $-0.54\pm0.77$ & $-0.11\pm0.15$ & $ 0.19\pm0.21$ & $46$ \\
   \hline
   (6) $M_{\rm lim}\rightarrow M^{*}_{B}(z)+2$ mag & 0.26  & $-0.42\pm0.83$ & $-0.11\pm0.15$ & $ 0.33\pm0.22$ & $76$ \\
   \hline
   (7)  $f_{b} \rightarrow 0.1$ $\star$            & 0.22  & $-0.12\pm0.77$ & $-0.03\pm0.16$ & $ 0.23\pm0.19$ & $46$ \\
   (8)  $f_{b} \rightarrow 0.0$\footnote{ No starburst (truncated star-formation).}
                                                   & 0.002 & $ 0.10\pm0.98$ & $ 0.02\pm0.19$ & $ 0.25\pm0.18$ & $99$ \\
   \hline
   (9)  $\tau \rightarrow 0.30$ Gyr $\star$        & 0.57  & $-0.03\pm0.76$ & $ 0.00\pm0.16$ & $ 0.27\pm0.17$ & $30$ \\
   \hline
   (10) Burst type : Exp.$\rightarrow$ Const.      & 0.11  & $-0.96\pm0.76$ & $-0.20\pm0.15$ & $ 0.13\pm0.20$ & $60$ \\
   \hline
   Observed\footnote{ The observed mean values with 1 $\sigma$ {\it errors}
   except for $f_{c,abs}$, where
   we give the accepted range.}
                                                   &       & $-0.18\pm0.33$ & $ 0.06\pm0.04$ & $ 0.20\pm0.05$ & $20$--$34$ \\
   \hline
   \end{tabular}
   \label{t_sim}
\end{minipage}
\end{table*}

\subsection{Evolution of star formation rate with redshift}

\begin{figure}
   \includegraphics[width=65mm,angle=-90]{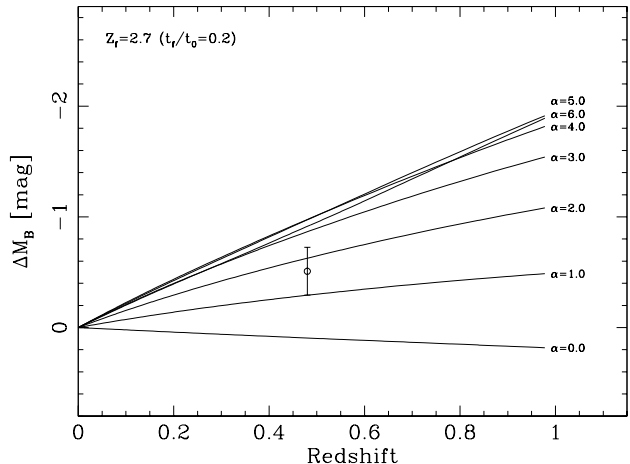}
   \caption{ The evolution of $B$-band 
             absolute magnitude with redshift for galaxies
	     with different star-formation histories. The lines
             show the model predictions with $SFR \propto (1+z)^{\alpha}$ for
	     different values of $\alpha$ 
             and solar metallicity. 
             The circle with the errorbar shows 
             the slope derived from Eq. (\ref{e_zev}) 
	     combined with the one derived by  B05B 
	     placed at the median redshift of the combined
             field TFR sample.
}
   \label{f_sfr_zev}
\end{figure}

The evolution of the $B$-band absolute magnitude with redshift  for the field
spirals in our sample (Eq. (\ref{e_zev}) and Fig.~\ref{f_zev}) can be
interpreted as due to the evolution of the star-formation rate (SFR) of these
galaxies. In this section, we parameterise the SFR evolution as a simple power
law of the  form $SFR \propto (1+z)^{\alpha}$,  and estimate the value of
$\alpha$ that best explains our result. In order to calculate the luminosity
evolution in the $B$-band  associated with the postulated SFR evolution we 
used Bruzual \& Charlot (2003) models with a Chabrier (2003) initial mass
function and solar metallicity. The results are shown in Fig.~\ref{f_sfr_zev}.
Changing the formation epoch of the galaxies in the range $1<z<5$  hardly
affects our estimated $\alpha$ for $\alpha\la2$. Thus, we arbitrarily set the
time at which spirals start forming stars at  20\% of the Universe's
present-day age  ($z_{\rm f}=2.7$ with the adopted `concordance' cosmology). 
Note that if the SFR had remained constant with time ($\alpha=0$), spiral 
galaxies would get slightly fainter with redshift in the $B$ band as 
the building up of their stellar populations with time slightly overcompensates
the average ageing of their stellar populations. Since our observations
indicate  that spiral galaxies get brighter with redshift, their SFR must
be higher at higher~$z$, i.e., $\alpha>0$. 

To compare our observational results with the model predictions, and to
increase the statistic accuracy, we compute the weighted-mean of the slopes in
Eq. (\ref{e_zev}) and in B05B ($-1.0\pm0.5$), and interpret it as the
representative of the evolution of field spiral  galaxies at the median
redshift of the VLT and Subary field galaxy samples.  We obtain $\Delta M_{B} =
-0.51\pm0.22$ mag at $z=0.48$, which is shown in Fig.~\ref{f_sfr_zev} as the
data-point with errorbars.  Taken at face value, our analysis yields
$\alpha=1.7\pm0.7$. 

This value could be an overestimate of the rate of evolution for two reasons.
First, in our analysis we have assumed constant (solar) metallicity for the
spiral galaxies. It is reasonable to expect that the average metallicity of
field spirals might have been lower in the past,  and, for a given stellar
mass, a lower metallicity stellar population is expected to be brighter in $B$
(see, e.g., Bruzual \& Charlot 2003). Thus,  part of the observed luminosity
evolution may be caused by a change in the average metallicity of the stellar
populations of the galaxies. However,  we expect this effect to be small
because the measured evolution in the average metallicity of field galaxies is
rather modest (cf. Kobulnicky et al.\ 2003; Lilly, Carollo \& Stockton 2003;
Kobulnicky \& Kewley 2004).  At the median redshift of our field galaxy sample
we expect the  average metallicity (at a given galaxy mass) to be only $\sim
0.07\,$dex lower than at $z=0$ (Kobulnicky \& Kewley 2004), which would have
negligible effect in our calculation. The second reason why 
we may have overestimated the rate of SFR evolution is that any unaccounted
for selection effects probably mean that we preferentially select 
intrinsically brighter galaxies at high-$z$, thus increasing the 
apparent luminosity evolution.

This rate of SFR evolution  is substantially shallower than the rate of
evolution of the SFR density of the universe indicated by studies based on the 
UV, H$\alpha$, far-infrared, and radio emission of galaxies, which all suggest
$\alpha \sim 3$--$4$ between $z=0$ and~$1$  (e.g., Hopkins 2004).
Hence, the rapid evolution in the cosmic  SFR
density  is not driven by  the evolution in the SFR
of individual bright ($M_B\la M_B^{*}$) spiral galaxies like the ones in our
sample, in agreement with the conclusions of B05B.

\section{Summary}

We have carried out MOS observation of 4 cluster fields using the FOCAS
spectrograph at Subaru,  and obtained spectra of 103 cluster and field spiral
galaxies with spectroscopic redshifts  in the range $0.06 \le z \le 1.20$. A
total of 77 galaxies show emission lines. Of these,  33 galaxies yielded
observed rotation curves of good enough quality to  determine secure rotation
velocities. Our sample reaches roughly $M_{B}^{*}$ at each redshift.

By comparing the rest-frame $B$-band  Tully-Fisher relation (TFR) of our
cluster  and field emission-line  spiral galaxies, we found no measurable
difference between the $B$-band luminosity at a given rotation velocity of 
both populations. This agrees with  the conclusions of Ziegler et al. (2003),
but disagrees with our own previous VLT results (Milvang-Jensen et al., 2003;
Bamford et al. 2005A).
We also find the rest-frame $B-V$ colour of the
cluster emission-line galaxies is marginally redder 
(by $0.06\pm0.04$) than
that of the field galaxies, providing little indication
that the cluster spirals are undergoing enhanced star formation.
On the other hand, we find that the fraction of spiral galaxies with
absorption-line spectra (i.e., no detectable emission lines) in the clusters is
larger than that in the field by a factor $\sim3$--$5$, in agreement with 
the results of Dressler et al. (1999). 
This implies that the cluster environment quenches star-formation on its
spiral galaxies, leading, perhaps, to the formation of S0s. 

To evaluate the significance of our results, we carried out simulations of  the
effects that changes in the star-formation history of in-falling spiral
galaxies would have on the observed properties of galaxy samples similar to
ours. Following the encounter of the in-falling galaxies with the cluster
environment, our models explore different star-formation scenarios, including 
truncation of the galaxies' star-formation on different timescales, possibly 
preceded by bursts of different strengths.  It turns out that even quite
drastic changes in the SFR of the galaxies may have quite modest effect on the
average luminosity and colour of a sample of bright cluster spiral galaxies
{\it selected to have ongoing star-formation\/} (i.e., having the strong
emission lines needed for rotation-curve measurement). However, the fraction of
spiral galaxies with absorption-line spectra (i.e., without current
star-formation) is very sensitive to the effect of star-formation truncation.  
Our Subaru observations favour models with relatively mild or absent initial
star-bursts, and relatively long star-formation timescales, while the VLT
results of Bamford et al. favour a more massive initial star-burst and a
shorter time-scale.  However, we estimate the probability that the observed
difference between our Subaru results and the VLT ones of Bamford et al.  
arises from ``unlucky'' statistic sampling of galaxy populations with 
intrinsically similar properties. We find that both Tully-Fisher relation
results are only different at the $\sim 1 \sigma$ level. This is due to  the
relatively small sample sizes and the large uncertainties  in the determination
of the individual TFR offsets. To definitively rule out the presence of a
starburst before the star-formation cessation would require
samples of $\sim50$--$100$ field and cluster spirals with reasonable
determinations of their rotation velocity. 

Finally, we find that the rest-frame absolute $B$-band magnitude  (at a fixed
rotation velocity) of the field galaxies in our sample shows an evolution of
$-1.30\pm1.04$ mag per unit redshift. By statistically-combining  our
luminosity evolution estimate with that of Bamford et al. (2005B;  $-1.0\pm0.5$
mag per unit redshift), and interpreting it as due to the increase with
redshift of the SFR of the galaxies, we estimate that   $SFR \propto
(1+z)^{1.7\pm0.7}$ for our field spirals.  This indicates that the average SFR
of bright ($M_B\la M_B^{*}$) disk galaxies evolves more slowly than  the
universal SFR density, suggesting that the evolution of the  global SFR
evolution is not dominated by bright star-forming disk galaxies, in agreement
with previous studies. 

\section*{Acknowledgments}

We would like to thank Yoshihiko Yamada and Masato Onodera for preparation
of the proposal for the observations.
We also thank the referee, Asmus B{\" o}hm, for his very useful comments. 
The work is partly supported by a Grant-in-Aid for Scientific Research
(No. 16540223) by the Japanese Ministry of Education, Culture,
Sports, Science and Technology.
This work is partly based on observations made with the NASA/ESA
{\it Hubble Space Telescope}, obtained from the data archive
at the Space Telescope Institute.
STScI is operated by the association of Universities for Research
in Astronomy, Inc. under the NASA contract NAS 5-26555.

\label{lastpage}
\end{document}